\pdfoutput=1
\newcommand*{\ATLASLATEXPATH}{latex/}
 \documentclass[atlasdraft=false, texlive=2016, USenglish, cernpreprint]{\ATLASLATEXPATH atlasdoc}

 \usepackage{\ATLASLATEXPATH atlaspackage}
 \usepackage{\ATLASLATEXPATH atlasbiblatex}

 \usepackage{\ATLASLATEXPATH atlasphysics}

 \graphicspath{{logos/}}

 \usepackage{rdphi-defs}

\AtlasTitle{Measurement of dijet azimuthal decorrelations 
            in $pp$ collisions at $\sqrt{s}=8\,$\TeV\/
            with the ATLAS detector  
            and determination of the strong coupling}

\AtlasAbstract{%
A measurement of the rapidity and transverse momentum dependence 
of dijet azimuthal decorrelations is presented, using the
quantity $\Rdphi$.
The quantity $\Rdphi$ specifies the fraction of the inclusive dijet
events in which the azimuthal opening angle of the two jets with the highest
transverse momenta is less than a given value of the parameter $\Dphimax$.
The quantity $\Rdphi$ is measured in proton--proton collisions at $\sqrt{s}=8\,$\TeV\/
as a function of the dijet rapidity interval,
the event total scalar transverse momentum, and $\Dphimax$.
The measurement uses an event sample corresponding to an integrated luminosity 
of $20.2\,$fb$^{-1}$ collected with the ATLAS detector at the
CERN Large Hadron Collider.
Predictions of a perturbative QCD calculation
at next-to-leading order in the strong coupling with corrections for
non-perturbative effects are compared to the data.
The theoretical predictions describe the data in the whole
kinematic region.
The data are used to determine the strong coupling $\as$ and to 
study its running for momentum transfers from 260\,\GeV\/ to above 1.6\,\TeV.
An analysis that combines data at all momentum transfers 
results in $\asmz = 0.1127^{+0.0063}_{-0.0027}$.
}

\AtlasRefCode{STDM-2012-19}

\PreprintIdNumber{CERN-EP-2017-282}

\AtlasJournal{Phys.\ Rev.\ D}

 \hypersetup{pdftitle={ATLAS document},pdfauthor={The ATLAS Collaboration}}

\begin{document}

 \maketitle

 \tableofcontents


\section{Introduction \label{sec:intro}}

In high-energy particle collisions,
measurements of the production rates
of hadronic jets with large transverse momentum \pt 
relative to the beam direction
can be employed to test the predictions
of perturbative quantum chromodynamics (pQCD).
The results can also be used
to determine the strong coupling $\as$,
and to test the pQCD predictions 
for the dependence of $\as$ on the
momentum transfer $Q$ (the ``running'' of $\as$) 
by the renormalization group equation 
(RGE)~\cite{Politzer:1973fx,Gross:1973ju}.
Previous tests of the RGE through $\as$ determinations
in hadronic final states
have been performed using data taken
in $ep$ collisions ($5<Q<60\,$\GeV)~\cite{Andreev:2016tgi,Andreev:2017vxu,Abramowicz:2012jz},
in $\epem$ annihilation ($10<Q<210\,$\GeV)~\cite{Bethke:2008hf,Dissertori:2009ik},
in $\ppbar$ collisions ($50<Q<400\,$\GeV)~\cite{Abazov:2009nc,Abazov:2012lua},
and in $pp$ collisions 
($130<Q<1400\,$\GeV)~\cite{Chatrchyan:2013txa,Khachatryan:2014waa,Khachatryan:2016mlc,CMS:2014mna,Aaboud:2017fml}.
The world average value is currently 
$\asmz = 0.1181 \pm 0.0011$~\cite{Olive:2016xmw}.

Recent $\as$ results from hadron collisions are limited by
theoretical uncertainties related to the scale dependence
of the fixed-order pQCD calculations.
The most precise $\asmz$ result from hadron collision data is
$\asmz = 0.1161 ^{+0.0041}_{-0.0048}$~\cite{Abazov:2009nc},
obtained from inclusive jet cross-section data,
using pQCD predictions beyond the next-to-leading order (NLO).
However, using the cross-section data in $\as$ determinations, 
the extracted $\as$ results are directly affected by our knowledge of the
parton distribution functions (PDFs) of the proton,
and their $Q$ dependence.
The PDF parameterizations depend on assumptions about $\as$ and the RGE
in the global data analyses in which they are determined.
Therefore, in determinations of $\as$ and its $Q$ dependence from 
cross-section data the RGE is already assumed in the inputs.
Such a conceptual limitation when using cross-section data
can largely be avoided by using ratios of multi-jet cross sections
in which PDFs cancel to some extent.
So far, the multi-jet cross-section ratios $\Rdr$~\cite{Abazov:2012lua} 
and $\Rtt$~\cite{Chatrchyan:2013txa} have been used
for $\as$ determinations at hadron colliders.
In this article, $\as$ is determined 
from dijet azimuthal decorrelations,
based on the multi-jet cross-section ratio $\Rdphi$~\cite{Wobisch:2012au}.
The RGE predictions are tested up to $Q = 1.675\,$\TeV.

The decorrelation of dijets in the azimuthal plane
has been the subject of a number of measurements
at the Fermilab Tevatron Collider~\cite{Abazov:2004hm}
and the CERN Large Hadron Collider 
(LHC)~\cite{Khachatryan:2011zj,daCosta:2011ni}.
The variable $\Dphi$ investigated in these analyses
is defined from the angles in the azimuthal plane
(the plane perpendicular to the beam direction)
$\phi_{1,2}$ of the two highest-\pt jets in the event
as $\Dphi  = | \phi_1 - \phi_2 |$.
In exclusive high-\pt dijet final states,
the two jets are correlated in the azimuthal plane
with $\Dphi = \pi$.
Deviations from this ($\Dphi < \pi$)
are due to additional activity in the final state,
as described in pQCD by processes of higher order in $\as$.
Due to kinematic constraints, the phase space in $2 \rightarrow 3$ 
processes is restricted to $\Dphi > 2\pi/3$~\cite{Wobisch:2015jea}
and lower $\Dphi$ values are only accessible in
$2 \rightarrow 4$ processes.
Measurements of dijet production with $2\pi/3 < \Dphi < \pi$ ($\Dphi < 2\pi/3$)
therefore test the pQCD matrix elements for three-jet (four-jet)
production.

The quantity $\Rdphi$ 
is defined as the fraction of all inclusive dijet events
in which $\Dphi$ is less than a specified value $\Dphimax$.
This quantity can be exploited
to extend the scope of the previous analyses
towards studies of the rapidity dependence
of dijet azimuthal decorrelations.
Since $\Rdphi$ is defined as a ratio of multi-jet cross sections
for which the PDFs cancel to a large extent,
it is well-suited for determinations of $\as$
and for studies of its running.

The quantity $\Rdphi$ has so far been measured
in $\ppbar$ collisions
at a center-of-mass energy of $\sqrt{s}=1.96\,$\TeV\/ at the
Fermilab Tevatron Collider~\cite{Abazov:2012jhu}.
This article presents the first measurement of $\Rdphi$
in $pp$ collisions,
based on data at $\sqrt{s}=8\,$\TeV\/ taken with the ATLAS detector
during 2012 at the LHC,
corresponding to an integrated luminosity 
of 20.2$\pm$0.4\,fb$^{-1}$~\cite{DAPR-2013-01}.
The data are corrected to ``particle level''~\cite{Buttar:2008jx},
and are used to extract $\as$ and to study its running
over a range of momentum transfers of $262 < Q < 1675\,$\GeV.

\section{Definition of $\Rdphi$ and the analysis phase space \label{sec:def}}

\begin{table}
\centering
\renewcommand{\arraystretch}{1.1}
\caption{\label{tab:ps} The values of the parameters and the requirements 
that define the analysis phase space for the inclusive dijet event sample.}
\begin{tabular}{lr}
\hline \hline
Variable & Value\\
\hline
$\ptmin$    & $100\,$\GeV    \\
$\yboost^\mathrm{max}$ & $0.5$  \\
$y^*_\mathrm{max}$ & $2.0$  \\
$\ptone/\Ht$    &   $>1/3$ \\
\hline \hline
\end{tabular}
\end{table}

The definitions of the quantity $\Rdphi$ and the 
choices of the variables that define the analysis phase space
are taken from the proposal in Ref.~\cite{Wobisch:2012au}.
Jets are defined by the \antikt\/ jet algorithm
as implemented in \textsc{fastjet}~\cite{Cacciari:2005hq,Cacciari:2008gp}.
The \antikt\/ jet algorithm is a successive recombination algorithm
in which particles are clustered into jets in the $E$-scheme
(i.e.\ the jet four-momentum is computed as the sum of the particle four-momenta).
The radius parameter is chosen to be $R=0.6$.
This is large enough for a jet to include a sufficient amount 
of soft and hard radiation around the jet axis, thereby improving the 
properties of pQCD calculations at fixed order in $\as$,
and it is small enough to avoid excessive contributions from 
the underlying event~\cite{Soyez:2010rg}.
An inclusive dijet event sample is extracted by selecting all events with 
two or 
more jets, where the two leading-\pt jets have $\pt > \ptmin$.
The dijet phase space is further specified in terms of the variables
$\yboost$ and $y^*$, computed from the rapidities, $y_1$ and $y_2$,
of the two leading-\pt jets 
as $\yboost = (y_1 + y_2) / 2$ and
$y^* = | y_1 - y_2| / 2$, respectively.\footnote{The ATLAS 
experiment uses a right-handed coordinate system, where
the origin is given by the nominal interaction point (IP) 
in the center of the detector.
The $x$-axis points from the IP to the center of the LHC ring, 
the $y$-axis points upward, and the $z$-axis along the proton beam direction.
Cylindrical coordinates ($r$, $\phi$) are used in the transverse plane, 
$\phi$ being the azimuthal angle around the beam pipe. 
The rapidity $y$ is defined as $y = \frac{1}{2} \ln \frac{E+p_z}{E-p_z}$,
and the pseudorapidity in terms of the polar angle $\theta$ 
as $\eta = - \ln \tan(\theta/2)$.
}
In $2\rightarrow 2$ processes, the variable 
$\yboost$ specifies the longitudinal boost
between the dijet and the proton--proton center-of-mass frames,
and $y^*$ (which is longitudinally boost-invariant) represents
the absolute value of the jet rapidities in the dijet center-of-mass frame.
The dijet phase space is restricted to
$|\yboost| < \yboost^\mathrm{max}$ and $y^* < y^*_\mathrm{max}$.
The variable \Ht is defined as the scalar sum of the jet \pt 
for all jets $i$ with $\pti > \ptmin$ and $|y_i - \yboost| < y^*_\mathrm{max}$.
Furthermore, the leading-\pt jet is required to have
$\ptone > \Ht/3$.
The values of the parameters $\ptmin$, $\yboost^\mathrm{max}$, 
and $y^*_\mathrm{max}$
ensure that jets are well-measured in the detector within $|y| < 2.5$
and that contributions from non-perturbative corrections
and pileup  (additional proton-proton interactions
within the same or nearby bunch crossings) are small.
The requirement $\ptone > \Ht/3$ ensures (for a given \Ht)
a well-defined minimum \ptone 
which allows single-jet triggers to be used in the measurement.
It also reduces the contributions from events with four or more jets,
and therefore pQCD corrections from higher orders in $\as$.
The values of all parameters are specified in Table~\ref{tab:ps}.
The quantity $\Rdphi$ is defined in this inclusive dijet event sample 
as the ratio
\begin{equation}
 \Rdphi(\Ht, y^*, \Dphimax) \, = \,
 \frac{
 \frac{d^2\sigma_\mathrm{dijet}(\Dphi < \Dphimax)}{d\Ht \, dy^*} }%
 {\frac{d^2\sigma_\mathrm{dijet}(\mbox{\footnotesize inclusive}) }{d\Ht \, dy^*}} \,,
 \label{eq:rdphi}
\end{equation}
where the denominator is the inclusive dijet cross section
in the phase space defined above,
in bins of the variables \Ht and $\ystar$.
The numerator is given by the subset of the denominator
for which $\Dphi$ of the two leading-\pt jets
obeys $\Dphi < \Dphimax$.
The measurement of the $y^*$ dependence of $\Rdphi$ allows a test 
of the rapidity dependence of the pQCD matrix elements.
The value of $\Dphimax$ is directly related to the hardness of the jet(s)
produced in addition to the two leading-\pt jets in the event.
The transverse momentum sum \Ht is one possible choice
that can be related to the scale 
at which $\as$ is probed.
The measurement is made as a function of \Ht in three different 
$\ystar$ regions and for four different values of $\Dphimax$ 
(see Table~\ref{tab:bins}).

\begin{table}
\centering
\renewcommand{\arraystretch}{1.1}
\caption{\label{tab:bins}
The \Ht, $y^*$, and $\Dphimax$ regions in which $\Rdphi(\Ht, y^*, \Dphimax)$
is measured.}
\begin{tabular}{lr}
\hline \hline
Quantity & Value\\
\hline
\Ht bin boundaries (in \TeV) & 0.45, 0.6, 0.75, 0.9, 1.1, \\
        & 1.4, 1.8, 2.2, 2.7, 4.0 \\
$y^*$ regions & $0.0$--$0.5$, $0.5$--$1.0$, $1.0$--$2.0$  \\
$\Dphimax$ values   & $7\pi/8$, $5\pi/6$, $3\pi/4$, $2\pi/3$    \\
\hline \hline
\end{tabular}
\end{table}

\section{Theoretical predictions \label{sec:theory}}

The theoretical predictions in this analysis
are obtained from perturbative calculations at fixed order in $\as$
with additional corrections for non-perturbative effects.

The pQCD calculations are carried out using
\nlojet~\cite{Nagy:2003tz,Nagy:2001fj}
interfaced to \fastnlo~\cite{Kluge:2006xs,Wobisch:2011ij}
based on the matrix elements for massless quarks
in the $\overline{\mathrm{MS}}$ scheme~\cite{Bardeen:1978yd}.
The renormalization and factorization scales are set
to $\mur = \muf = \mu_0$ with $\mu_0 = \Ht/2$.
In inclusive dijet production at leading order (LO) in pQCD 
this choice is equivalent to other common choices:
$\mu_0 = \overline{\pt} = (\ptone + \pttwo)/2$ and $\mu_0 = \ptone$.
The evolution of $\as$ is computed using the 
numerical solution of the next-to-leading-logarithmic
(2-loop) approximation of the RGE.

The pQCD predictions for the ratio $\Rdphi$ are obtained
from the ratio of the cross sections in the numerator
and denominator in Eq. (\ref{eq:rdphi}),
computed to the same relative order (both either to NLO or to LO).
The pQCD predictions for the cross section in the denominator 
by \nlojet\/ are available up to NLO.
For $\Dphimax = 7\pi/8, 5\pi/6, 3\pi/4$ ($2\pi/3$) the numerator 
is a three-jet (four-jet) quantity for which the pQCD predictions 
in \nlojet\/ are available up to NLO (LO)~\cite{Wobisch:2015jea}.

The PDFs are taken from the global analyses 
MMHT2014 (NLO)~\cite{Harland-Lang:2014zoa,Harland-Lang:2015nxa},
CT14 (NLO)~\cite{Dulat:2015mca}, and NNPDFv2.3 
(NLO)~\cite{Ball:2011mu}.\footnote{The NNPDFv3.0 PDFs~\cite{Ball:2014uwa} 
are available only for
a rather limited $\asmz$ range ($0.115$--$0.121$); 
therefore, the older NNPDFv2.3 results are employed.}
For additional studies, the PDF sets
ABMP16 (NNLO)~\cite{Alekhin:2017kpj}\footnote{The ABMP16 analysis 
does not provide NLO PDF sets for a series of $\asmz$ values;
their NNLO PDF sets are therefore used.} 
and HERAPDF~2.0 (NLO)~\cite{Abramowicz:2015mha}
are used, 
which were obtained using data from selected processes only.
All of these PDF sets were obtained for a series of discrete 
$\asmz$ values, in increments of $\Delta\asmz = 0.001$
(or $\Delta\asmz = 0.002$ for NNPDFv2.3).
In all calculations in this article, the PDF sets
are consistently chosen to correspond to the value of $\asmz$
used in the matrix elements.
The extraction of $\as$ from the experimental $\Rdphi$ data requires
a continuous dependence of the pQCD calculations on $\asmz$.
This is obtained by cubic interpolation (linear extrapolation)
for $\asmz$ values inside (outside) the ranges 
provided by the PDF sets.
The central predictions that are compared to the data use
$\asmz = 0.118$, which is close to the current world average,
and the MMHT2014 PDFs.
The MMHT2014 PDFs also provide the largest 
range of $\asmz$ values ($0.108 \le \asmz \le 0.128$).
For these reasons, the MMHT2014 PDFs are used to obtain the central results
in the $\as$ determinations.

The uncertainties of the perturbative calculation are estimated
from the scale dependence 
(as an estimate of missing higher-order pQCD corrections)
and the PDF uncertainties.
The former is evaluated from independent variations of $\mur$ and $\muf$
between $\mu_0/2$ and $2 \mu_0$
(with the restriction $0.5 \le \mur / \muf \le 2.0$).
The PDF-induced uncertainty is computed by propagating the 
MMHT2014 PDF uncertainties.
In addition, a ``PDF set'' uncertainty is included as the
envelope of the differences of the results 
obtained with CT14, NNPDFv2.3, ABMP16, and HERAPDF~2.0, relative to 
those obtained with MMHT2014.

The pQCD predictions based on matrix elements for massless quarks
also depend on the number of quark flavors, 
in gluon splitting ($g \rightarrow q\bar{q}$), $n_\textrm{f}$,
which affects the tree-level matrix elements 
and their real and virtual corrections,
as well as the RGE predictions 
and the PDFs obtained from global data analyses.
The central results in this analysis are obtained for a
consistent choice $n_\textrm{f}=5$ in all of these contributions.
Studies of the effects of using $n_\textrm{f}=6$ in the matrix elements and 
the RGE, as documented in Appendix~\ref{sec:top},
show that the corresponding effects for $\Rdphi$ 
are between $-1\%$ and $+2\%$
over the whole kinematic range of this measurement.
Appendix~\ref{sec:top} also includes a study of the contributions
from the $t\bar{t}$ production process,
concluding that the effects on $\Rdphi$ are less than 0.5\% 
over the whole analysis phase space.

The corrections due to non-perturbative effects, related to hadronization
and the underlying event, were obtained in Ref.~\cite{Wobisch:2012au},
using the event generators \pythia\/ 6.426~\cite{Sjostrand:2000wi} 
and \herwig\/ 6.520~\cite{Corcella:2000bw,Corcella:2002jc}.
An estimate of the model uncertainty is obtained from a study of the
dependence on the generator's parameter settings (tunes),
based on the \pythia\/ tunes AMBT1~\cite{ATLAS-CONF-2010-031},
DW~\cite{Albrow:2006rt}, A~\cite{tuneA}, and S-Global~\cite{Schulz:2011qy},
which differ in the parameter settings and the
implementations of the parton-shower and underlying-event models.
All model predictions for the total non-perturbative corrections
lie below 2\% (4\%) for $\Dphimax = 7\pi/8$ and $5\pi/6$
($\Dphimax = 3\pi/4$ and $2\pi/3$),
and the different models agree within 2\% (5\%) 
for $\Dphimax = 7\pi/8$ and $5\pi/6$
($\Dphimax = 3\pi/4$ and $2\pi/3$).

For this analysis, the central results are taken to be 
the average values obtained from \pythia\/ with tunes AMBT1 and DW.
The corresponding uncertainty is taken to be half of the difference 
(the numerical values are provided in Ref.~\cite{suppl}).
The results obtained with \pythia\ tunes A and S-Global as well as \herwig\ 
are used to study systematic uncertainties.

\section{ATLAS detector \label{sec:atlas}}

ATLAS is a general-purpose detector consisting of an inner tracking detector,
a calorimeter system, a muon spectrometer, and magnet systems.
A detailed description of the ATLAS detector is given
in Ref.~\cite{Aad:2008zzm}.
The main components used in the $\Rdphi$ measurement
are the inner detector, the calorimeters, and the trigger system.

The position of the $pp$ interaction is determined from
charged-particle tracks reconstructed in the inner detector,
located inside a superconducting solenoid that provides a 
2~T axial magnetic field.
The inner detector, covering the region $|\eta|<2.5$,
consists of layers of silicon pixels,
silicon microstrips, and transition radiation tracking detectors.

Jet energies and directions are measured in the three electromagnetic and
four hadronic calorimeters with a coverage of $|\eta|<4.9$.
The electromagnetic liquid argon (LAr) calorimeters
cover $|\eta| < 1.475$ (barrel), $1.375 < |\eta| < 3.2$ (endcap),
and $3.1 < |\eta| < 4.9$ (forward).
The regions $|\eta| < 0.8$ (barrel) and $0.8 < |\eta| < 1.7$ (extended barrel)
are covered by scintillator/steel sampling hadronic calorimeters,
while the regions $1.5 < |\eta| < 3.2$ and $3.1 < |\eta| < 4.9$
are covered by the hadronic endcap with LAr/Cu calorimeter modules,
and the hadronic forward calorimeter with LAr/W modules.

During 2012, for $pp$ collisions, 
the ATLAS trigger system was divided into three levels, 
labeled L1, L2, and the Event Filter (EF)~\cite{Aad:2012xs,ATLAS:2016qun}. 
The L1 trigger is hardware-based, while L2 and EF are software-based and 
impose increasingly refined selections designed to identify events of interest. 
The jet trigger identifies electromagnetically and hadronically interacting 
particles by reconstructing the energy deposited in the calorimeters. 
The L1 jet trigger uses a sliding window of
$\Delta\eta \times \Delta\phi = 0.8 \times 0.8$ 
to find jets
and requires these to have transverse energies \et above a given threshold, 
measured at the electromagnetic scale. 
Jets triggered by L1 are passed to the L2 jet trigger, which reconstructs jets 
in the same region using a simple cone jet algorithm with a cone size of 
$0.4$ in ($\eta$, $\phi$) space.
Events are accepted if a L2 jet is above a given \et threshold. 
In events which pass L2, a full event reconstruction is performed by the EF. 
The jet EF constructs topological clusters~\cite{Aad:2016upy} 
from which jets are then formed, 
using the \antikt\/ jet algorithm with a radius parameter of $R = 0.4$. 
These jets are then calibrated to the hadronic scale.
Events for this analysis are collected either with
single-jet triggers with different minimum \et requirements
or with multi-jet triggers based on a single high-\et jet 
plus some amount of \Ht (the scalar \et sum)
of the multi-jet system.
The trigger efficiencies are determined relative to
fully efficient reference triggers,
and each trigger is used above an \Ht threshold 
where it is more than 98\% efficient.
The triggers used for the different \Ht regions 
in the offline analysis
are listed in Table~\ref{tab:trigger}.

Single-jet triggers select events if any jet with $|\eta| <3.2$ is above the 
\et thresholds at L1, L2, and the EF. 
Due to their high rates, the single-jet triggers studied 
are highly prescaled during data-taking. 
Multi-jet triggers select events if an appropriate high-\et jet 
is identified and the \Ht value, summed over all jets at the EF with $|\eta| < 3.2$ 
and $\et>45$~\GeV, is above a given threshold. 
The additional \Ht requirement significantly reduces the selected event 
rate, and lower prescales can be applied. 
The integrated luminosity of the data sample collected with the 
highest threshold triggers is 20.2$\pm$0.4\,fb$^{-1}$.

\begin{table}
\centering
\renewcommand{\arraystretch}{1.1}
\caption{\label{tab:trigger}
The triggers used to select the multi-jet events
in the different \Ht ranges in the offline analysis,
and the corresponding integrated luminosities.}
\begin{tabular}{ccr}
\hline \hline
\Ht range [\GeV] &  Trigger type & Integrated luminosity [pb$^{-1}$]  \\
\hline
450--600  & single-jet & 9.6 $\pm$ 0.2 \\
600--750 & single-jet  & 36 $\pm$ 1 \\
750--900 & multi-jet & 546 $\pm$ 11 \\
$>$900 & multi-jet & (20.2 $\pm$ 0.4) $\cdot$ 10$^{3}$ \\ 
\hline \hline
\end{tabular}
\end{table}

The detector response for the measured quantities
is determined using a detailed simulation of the ATLAS detector
in \GEANT~4~\cite{Agostinelli:2002hh,Aad:2010ah}.
The particle-level events, subjected to the detector simulation,
were produced by the \pythia\/ event generator~\cite{Sjostrand:2007gs} 
(version 8.160) with CT10 PDFs.
The \pythia\/ parameters were set according to the 
AU2 tune~\cite{ATL-PHYS-PUB-2012-003}.
The ``particle-level'' jets are defined
based on the four-momenta of the generated stable particles 
(as recommended in Ref.~\cite{Buttar:2008jx},
with a proper lifetime $\tau$ satisfying $c\tau > 10\,$mm, including
muons and neutrinos from hadron decays).
The ``detector-level'' jets are defined based on the
four-momenta of the simulated detector objects.

\section{Measurement procedure \label{sec:measure}}

The inclusive dijet events used for the measurement of $\Rdphi$ were
collected between April and December 2012 by the ATLAS detector
in proton--proton collisions at $\sqrt{s} = 8\,$\TeV. 
All events used in this measurement are required 
to satisfy data-quality criteria 
which include stable beam conditions and stable operation of the tracking systems, 
calorimeters, solenoid, and trigger system. 
Events that pass the trigger selections described above are included 
in the sample if they contain at least one primary collision vertex 
with at least two associated tracks with $\pt > 400\,$\MeV, 
in order to reject contributions due to cosmic-ray events and beam background.
The primary vertex with highest $\sum \pt^2$ of associated tracks is taken as 
the event vertex.

Jets are reconstructed offline using the \antikt\/ jet algorithm 
with a radius parameter $R=0.6$. 
Input to the jet algorithm consists of locally calibrated three-dimensional 
topological clusters~\cite{Aad:2016upy}
formed from sums of calorimeter cell energies, 
corrected for local calorimeter response, dead material, 
and out-of-cluster losses for pions.
The jets are further corrected for pileup contributions and then 
calibrated to the hadronic scale, 
as detailed in the following. 
The pileup correction is applied to account for the effects on the jet response from 
additional interactions within the same proton bunch crossing (``in-time pileup'') and 
from interactions in bunch crossings preceding or following the one of interest 
(``out-of-time pileup''). 
Energy is subtracted from each jet, based upon the energy density in the event  
and the measured area of the jet~\cite{Cacciari:2007fd}.
The jet energy is then adjusted by a small residual correction
depending on the average pileup conditions for the event.  
This calibration restores the calorimeter energy scale, on average, to a reference 
point where pileup is not present~\cite{PERF-2014-03}.
Jets are then calibrated using an energy- and $\eta$-dependent correction 
to the hadronic scale with constants derived 
from data and Monte Carlo samples of jets produced in multi-jet processes.
A residual calibration, based on a combination of several in situ 
techniques, is applied to take into account 
differences between data and Monte Carlo simulation.
In the central region of the detector, the uncertainty in the 
jet energy calibration
is derived from the transverse momentum balance in $Z$+jet, $\gamma$+jet
or multi-jet events measured in situ,
by propagating the known uncertainties of the energies of the reference objects 
to the jet energies.
The energy uncertainties for the central region are then propagated 
to the forward region by studying the
transverse momentum balance in dijet events
with one central and one forward jet~\cite{ATLAS-CONF-2015-037}.
The energy calibration uncertainty in the high-\pt range 
is estimated using the in situ measurement of
the response to single isolated hadrons~\cite{Aaboud:2016hwh}.
The jet energy calibration's total uncertainty is decomposed 
into 57 uncorrelated contributions, of which each is fully correlated
in \pt.
The corresponding uncertainty in jet \pt is between $1\%$ and $4\%$ 
in the central region ($|\eta| < 1.8$), and increases to $5\%$ in the
forward region ($1.8 < |\eta| < 4.5$).

The jet energy resolution has been measured in the data using the
bisector method in dijet events~\cite{ATLAS-CONF-2015-017,Aad:2012ag,ATLAS-CONF-2015-057}
and the Monte Carlo simulation is seen to be in good agreement with the data.
The uncertainty in the jet energy resolution is affected by 
selection parameters for jets, 
such as the amount of nearby jet activity, 
and depends on the $\eta$ and \pt values of the jets.
Further details about the determinations of the jet energy scale and resolution
are given 
in Refs.~\cite{ATLAS:2015oia,ATLAS-CONF-2015-017,Aaboud:2016hwh}.

The angular resolution of jets is obtained in the Monte Carlo simulation
by matching particle-level jets with detector-level jets,
when their distance in $\DR = \sqrt{(\Delta y^2 + \Delta \phi^2)}$ 
is smaller than the jet radius parameter.
The jet $\eta$ and $\phi$ resolutions are obtained
from a Gaussian fit to the distributions of the difference
between the detector-level and particle-level values 
of the corresponding quantity.
The difference between the angular resolutions determined from
different Monte Carlo simulations is taken as a systematic uncertainty
for the measurement result, 
which is about $10$--$15\%$ for $\pt < 150\,$\GeV\/ 
and decreases to about $1\%$ for $\pt > 400\,$\GeV. 
The bias in jet $\eta$ and $\phi$ is found to be negligible.

All jets within the whole detector acceptance, $|\eta| < 4.9$, 
are considered in the analysis.
Data-quality requirements are applied to each reconstructed jet
according to its properties,
to reject spurious jets not originating from hard-scattering events. 
In each \Ht bin, events from a single trigger are used and
the same trigger is used for the numerator and the denominator
of $\Rdphi$.
In order to test the stability of the measurement results, 
the event sample is divided into subsamples with different pileup conditions.
The $\Rdphi$ results for different pileup conditions
are compatible within the statistical uncertainties
without any systematic trends.
The measurement is also tested for variations resulting from loosening 
the requirements on
the event- and jet-data-quality conditions, and the observed variations 
are also consistent within the statistical uncertainties.

The distributions of $\Rdphi(\Ht, y^*, \Dphimax)$
are corrected for experimental effects, including detector resolutions
and inefficiencies, using the simulation.
To ensure that the simulation describes all relevant distributions,
including the \pt and $y$ distributions
of the jets, the generated events are reweighted, 
based on the properties of the 
generated jets, to match these distributions in data, and to match
the \Ht dependence of the observed inclusive dijet cross section
as well as the $\Rdphi$ distributions and their \Ht dependence.
To minimize migrations between \Ht bins due to resolution 
effects, the bin widths are chosen to be larger than the detector resolution.
The bin purities, defined as the fraction of all 
reconstructed events that are generated in the same bin, 
are $65$--$85\%$ for $\Dphimax = 7\pi/8$ and $5\pi/6$,
and $50$--$75\%$ for  $\Dphimax = 3\pi/4$ and $2\pi/3$.
The bin efficiencies, defined as the fraction of all 
generated events that are reconstructed in the same bin, 
have values in the same ranges as the bin purities.
The corrections are obtained bin by bin from the generated 
\pythia\ events as the ratio of the $\Rdphi$ results for 
the particle-level jets and the detector-level jets.
These corrections are typically between 0\% and 3\%, and never 
outside the range from $-10\%$ to $+10\%$.
Uncertainties in these corrections due to the modeling of 
the migrations by the simulation
are estimated from the changes of the correction factors
when varying the reweighting function.
In most parts of the phase space, these uncertainties are below 1\%.
The results from the bin-by-bin correction procedure were compared
to the results when using a Bayesian iterative unfolding 
procedure~\cite{DAgostini:1994fjx},
and the two results agree within their statistical uncertainties.

\begin{figure}
  \centering
  \includegraphics{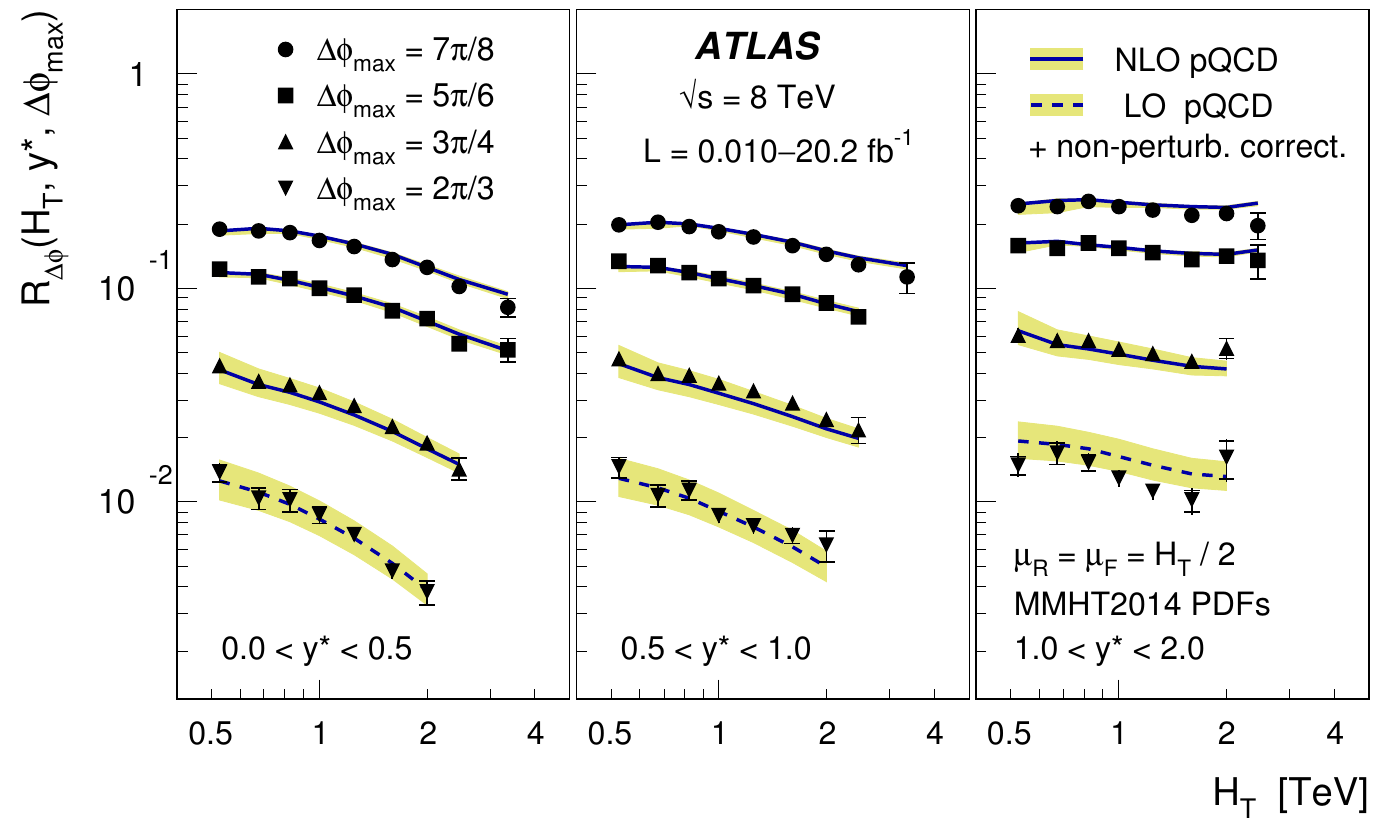}
  \caption{The measurement of $\Rdphi(\Ht, y^*, \Dphimax)$ as a function of \Ht
         in three regions of $\ystar$ and for four choices of $\Dphimax$.
   The inner error bars indicate the statistical uncertainties,
   and the sum in quadrature of statistical and systematic uncertainties
   is displayed by the total error bars.
   The theoretical predictions, based on pQCD at NLO
   (for  $\Dphimax = 7\pi/8$, $5\pi/6$, and $3\pi/4$)
   and LO (for $\Dphimax = 2\pi/3$) are shown as solid and dashed lines, 
   respectively.
   The shaded bands display the PDF uncertainties and the scale dependence,
   added in quadrature.}
  \label{fig:rdphi1}
\end{figure}

The uncertainties of the $\Rdphi$ measurements include
two sources of statistical uncertainty and 62 sources of
systematic uncertainty.
The statistical uncertainties arise from the data and from the correction factors.
The systematic uncertainties are from the correction factors 
(two independent sources, related to variations of the reweighting 
of the generated events),
the jet energy calibration (57 independent sources), 
the jet energy resolution, 
and the jet $\eta$ and $\phi$ resolutions.
To avoid double counting of statistical fluctuations, the \Ht dependence 
of the uncertainty distributions is smoothed by fitting
either linear or quadratic  functions in $\log(\Ht/\mbox{\GeV})$.
From all 62 sources of experimental correlated uncertainties,
the dominant systematic uncertainties are due to the jet energy calibration.
For $\Dphimax = 7\pi/8$ and $5\pi/6$
the jet energy calibration uncertainties are typically 
between 1.0\% and 1.5\% and always less than 3.1\%.
For smaller values of $\Dphimax$ they
can be as large as 4\% (for $\Dphimax = 3\pi/4$) 
or 9\% (for $\Dphimax = 2\pi/3$).
A comprehensive documentation of the measurement results, 
including the individual contributions 
due to all independent sources of uncertainty, 
is provided in Ref.~\cite{suppl}.

\begin{figure}
  \centering
  \includegraphics{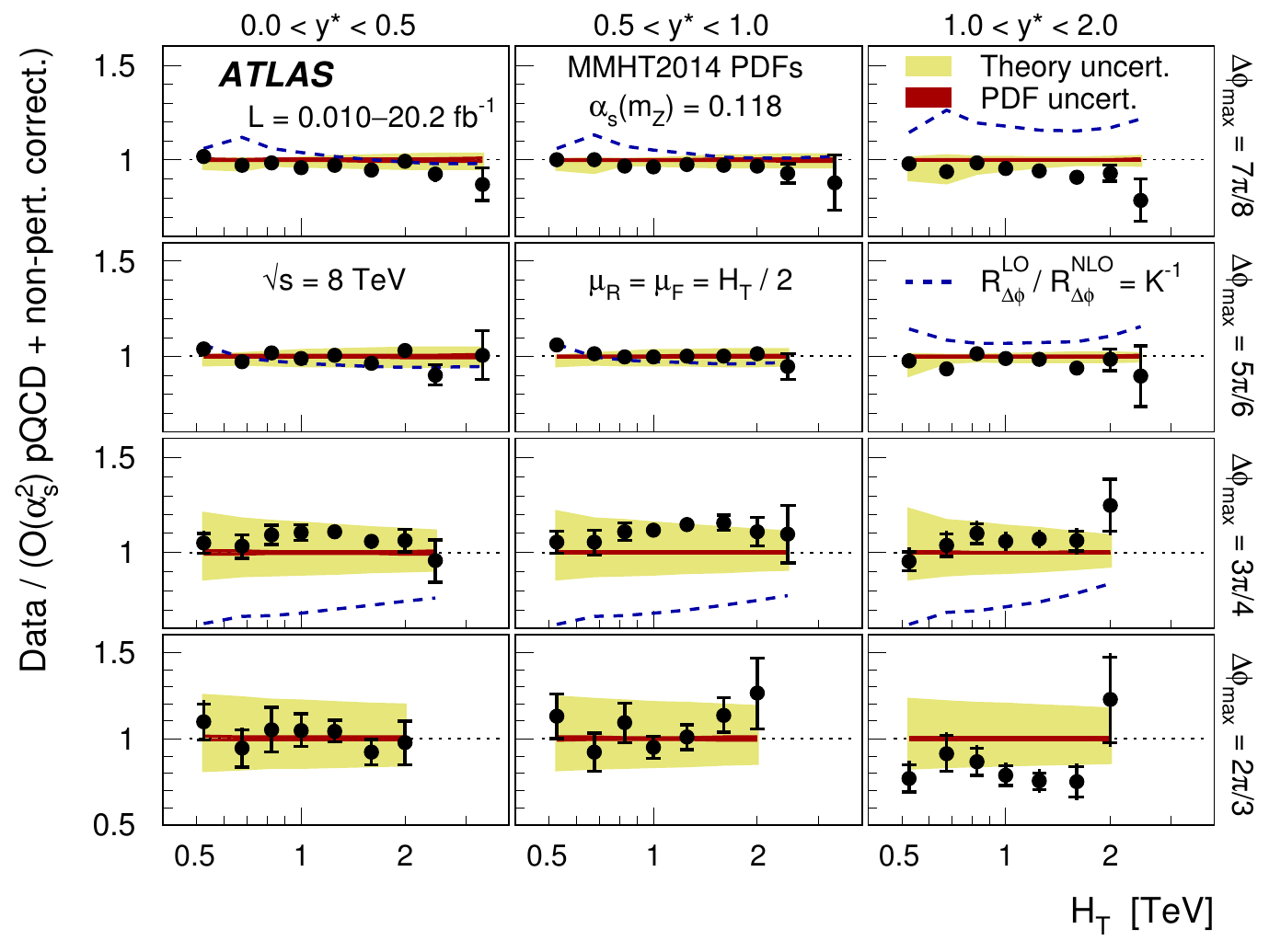}
  \caption{The ratios of the $\Rdphi$ measurements and the 
       theoretical predictions 
       obtained for MMHT2014 PDFs and $\asmz = 0.118$.
       The ratios are shown as a function of \Ht, 
       in different regions of $y^*$ (columns) and for different $\Dphimax$ 
       (rows).
      The inner error bars indicate the statistical uncertainties
      and the sum in quadrature of statistical and systematic uncertainties
      is displayed by the total error bars.
      The theoretical uncertainty is the sum in quadrature 
      of the uncertainties 
      due to the PDFs and the scale dependence.
      The inverse of the NLO $K$-factor is indicated by the dashed line.
  \label{fig:rdphi2}}
\end{figure}

\section{Measurement results \label{sec:expresults}}

The measurement results for $\Rdphi(\Ht, \ystar, \Dphimax)$ 
are corrected to the particle level 
and presented as a function of \Ht, in different regions of $\ystar$
and for different $\Dphimax$ requirements.
The results are listed in Appendix~\ref{sec:datatable} in 
Tables~\ref{tab:data1}--\ref{tab:data4},
and displayed in Figure~\ref{fig:rdphi1},
at the arithmetic center of the \Ht bins.
At fixed ($y^*$, $\Dphimax$),
$\Rdphi(\Ht, \ystar, \Dphimax)$ decreases with increasing \Ht 
and increases with increasing $y^*$ at fixed (\Ht, $\Dphimax$).
At fixed (\Ht, $y^*$), $\Rdphi$ decreases with decreasing $\Dphimax$.

Theoretical predictions based on NLO pQCD 
(for $\Dphimax = 7\pi/8$, $5\pi/6$, and $3\pi/4$)
or LO (for $\Dphimax= 2\pi/3$)
with corrections for non-perturbative effects,
as described in Section~\ref{sec:theory},
are compared to the data.
The ratios of data to the theoretical predictions are displayed in 
Figure~\ref{fig:rdphi2}.
To provide further information about the convergence of the pQCD 
calculation, the inverse of the NLO $K$-factors are also shown
(defined as the ratio of predictions for $\Rdphi$ at NLO and LO,
$K = \Rdphi^\mathrm{NLO}/\Rdphi^\mathrm{LO}$).
In all kinematical regions, the data are described by 
the theoretical predictions, even for $\Dphimax = 2\pi/3$,
where the predictions are only based on LO pQCD
and have uncertainties of about $20\%$ 
(dominated by the dependence on $\mur$ and $\muf$).
The data for $\Dphimax = 7\pi/8$ and $5\pi/6$ 
allow the most stringent tests of the theoretical predictions,
since for these $\Dphimax$ values the theoretical uncertainties 
are typically less than $\pm5\%$.

\section{Selection of data points for the $\as$ extraction \label{sec:criteria}}

The extraction of $\as(Q)$ at different scales $Q = \Ht/2$
is based on a combination of data points 
in different kinematic regions of $\ystar$ and $\Dphimax$,
with the same \Ht.
The data points are chosen according to the following criteria.

\begin{enumerate}

\item
  Data points are used only from kinematic regions in which the 
  pQCD predictions appear to be most reliable, 
  as judged by the renormalization
  and factorization scale dependence, and by the NLO $K$-factors.

\item
  For simplicity, data points are only combined in the $\as$ extraction
  if they are statistically independent,
  i.e.\ if their accessible phase space does not overlap.

\item 
  The preferred data points are those for which the
  cancellation of the PDFs between the numerator and the denominator
  in $\Rdphi$ is largest.
 
\item
  The experimental uncertainty at large \Ht is limited by the sample size.
  If the above criteria give equal preference to two or more data sets
  with overlapping phase space, the data points with smaller
  statistical uncertainties are used to test the RGE at the largest possible 
  momentum transfers with the highest precision.
 
\end{enumerate}

Based on criterion (1), the data points obtained for $\Dphimax = 2\pi/3$ 
are excluded, as the pQCD predictions in \nlojet\/ are only available at LO. 
Furthermore, it is observed that the points for $\Dphimax = 3\pi/4$ 
have a large scale dependence, which is typically between $+15\%$ and $-10\%$.
For the remaining data points with
$\Dphimax = 7\pi/8$ and $5\pi/6$ at larger $y^*$ ($1 < y^*< 2$), the NLO
corrections are negative and 
(with a size of $5$--$23\%$) larger than those
at smaller $y^*$, indicating
potentially larger corrections from not yet calculated higher orders. 
The conclusion from criterion (1) is therefore that
the pQCD predictions are most reliable in the four kinematic regions 
$0 < y^* < 0.5$ and $0.5 < y^* < 1$, 
for $\Dphimax = 7\pi/8$ and $\Dphimax = 5\pi/6$, where the
NLO $K$-factors are typically within $\pm5\%$ of unity.

The requirement of statistically independent data points according 
to criterion (2) means that the data points from different $y^*$ regions 
can be combined,
but not those with different $\Dphimax$. 
The choice whether to use the data with $\Dphimax = 7\pi/8$ or
$5\pi/6$ (in either case combining the data for 
$0 < y^* < 0.5$ and $0.5 < y^* < 1$)
is therefore based on criteria (3) and (4).

The cancellation of the PDFs, as addressed in criterion (3),
is largest 
for those data points for which the phase space of the numerator
in Eq.~(\ref{eq:rdphi}) is closest to that of the denominator.
Since the numerator of $\Rdphi$ is a subset of the denominator,
this applies more to the data at larger values of $\Dphimax$.
For those points, the fractional contributions from different 
partonic subprocesses ($gg \rightarrow$ jets, $gq \rightarrow$ jets, 
$qq \rightarrow$ jets), and the ranges in the accessible
proton momentum fraction $x$
are more similar for the numerator and denominator,
resulting in a larger cancellation of PDFs in $\Rdphi$.
This argument, based on the third criterion, leads to the same
conclusion as the suggestion of criterion (4), to use
the data set with smallest statistical uncertainty.

Based on the four criteria, $\as$ is therefore extracted combining
the data points in the rapidity regions $0 < y^* < 0.5$ and $0.5 < y^* < 1$ 
for $\Dphimax = 7\pi/8$.
Extractions of $\as$ from the data points in other
kinematical regions in $y^*$ and $\Dphimax$ 
are used to investigate the dependence of the final results on 
those choices.

\begin{figure}
\centering
\includegraphics[scale=1.1]{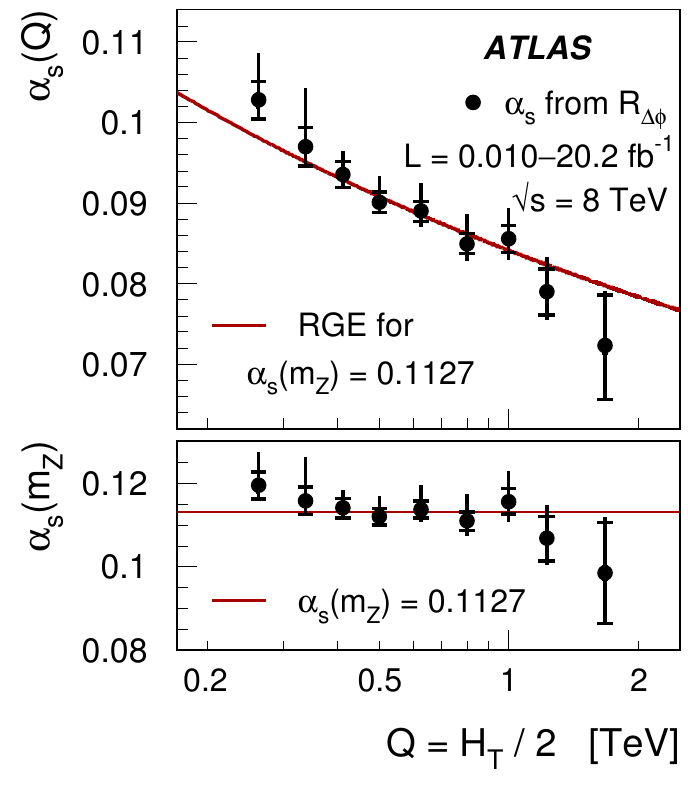}
\caption{\label{fig:alphas1} The $\as$ results determined from the 
   $\Rdphi$ data for $\Dphimax = 7\pi/8$ in the $y^*$ regions 
   $0<y^*<0.5$ and $0.5 < y^* < 1.0$ in the range of $262 < Q < 1675\,$\GeV.
   The inner error bars indicate the experimental uncertainties
   and the sum in quadrature of experimental and theoretical uncertainties
   is displayed by the total error bars.
   The $\asq$ results (top) are displayed together
   with the prediction of the RGE for 
   the $\asmz$ result obtained in this analysis.
   The individual $\asq$ values are then evolved to $Q=\mZ$ (bottom).
}
\end{figure}

\section{Determination of $\as$ \label{sec:alphas}}

The $\Rdphi$ measurements in the selected kinematic regions
are used to determine
$\as$ and to test the QCD predictions for its running as a function
of the scale $Q = \Ht/2$.
The $\as$ results are extracted by using \textsc{minuit}~\cite{James:1975dr},
to minimize the $\chi^2$ function specified in Appendix~\ref{sec:chi2}.
In this approach, the experimental and theoretical uncertainties
that are correlated between all data points are treated
in the Hessian method~\cite{Alekhin:2005dx}
by including a nuisance parameter for each uncertainty source,
as described in Appendix~\ref{sec:chi2}.
The only exceptions are the uncertainties due to the PDF set 
and the $\murf$ dependence of the pQCD calculation.
These uncertainties are determined from the variations of the $\as$ results,
when repeating the $\as$ extractions for different PDF sets
and for variations of the scales $\murf$ as described 
in Section~\ref{sec:theory}.

Results of $\asq$ (with $Q=\Ht/2$, taken at the arithmetic centers of 
the \Ht bins) are determined from the 
$\Rdphi$ data for $\Dphimax = 7\pi/8$, combining
the data points in the two $y^*$ regions of $0<y^* < 0.5$ and $0.5<y^* < 1.0$. 
Nine $\as(Q)$ values are determined in the range $262 < Q \le 1675\,$\GeV.
A single $\chi^2$ minimization provides the uncertainties
due to the statistical uncertainties, the experimental correlated uncertainties,
the uncertainties due to the non-perturbative corrections,
and the MMHT2014 PDF uncertainty.
Separate $\chi^2$ minimizations are made for variations of
$\mur$ and $\muf$ (in the ranges described in Section~\ref{sec:theory}),
and also for the CT14, NNPDFv2.3, ABMP16, and HERAPDF~2.0 PDF sets.
The largest individual variations are used to quantify the uncertainty due 
to the scale dependence and the PDF set, respectively.
The so-defined PDF set uncertainty may partially double count
some of the uncertainties already taken into account by the
MMHT2014 PDF uncertainties, 
but it may also include some additional systematic uncertainties 
due to different approaches in the PDF determinations. 
The $\asq$ results are displayed in Figure~\ref{fig:alphas1}
and listed in Table~\ref{tab:asq}.

\begin{table*}[t]
\centering
\renewcommand{\arraystretch}{1.1}
\caption{\label{tab:asq}
The results for $\asq$ determined from the $\Rdphi$ data
for $\Dphimax = 7\pi/8$ with $0<\ystar<0.5$ and $0.5<\ystar<1.0$.
All uncertainties have been multiplied by a factor of $10^3$.}
\begin{tabular}{c@{\hskip 10pt}c@{\hskip 10pt}ccccccc}
\hline \hline
$Q$  & $\asq$  &
Total  &
Stat. & Exp. & Non-perturb. & MMHT2014 & PDF & $\murf$ \\
 \multicolumn{1}{c}{[\GeV]}    & & uncert. & & correlated &
corrections & uncertainty & set & variation \\
\hline
   262.5 &
   0.1029 & $^{+ 6.0}_{- 2.8}$ & $\pm 1.6$
  & $^{+ 1.6}_{- 1.7}$ & $^{+ 0.4}_{- 0.4}$ & $^{+ 0.4}_{- 0.4}$ 
   & $^{+ 1.4}_{- 0.9}$ & $^{+ 5.3}_{- 0.2}$ 
 \\[2pt]
   337.5 &
    0.0970 & $^{+ 8.0}_{- 2.6}$ & $\pm 1.8$
    & $^{+ 1.5}_{- 1.5}$ & $^{+ 0.4}_{- 0.4}$ & $^{+ 0.3}_{- 0.3}$ 
   & $^{+ 3.0}_{- 0.5}$  & $^{+ 7.0}_{- 0.7}$ 
 \\[2pt]
   412.5 &
     0.0936 & $^{+ 4.0}_{- 2.2}$ & $\pm 0.9$
    & $^{+ 1.3}_{- 1.3}$ & $^{+ 0.3}_{- 0.3}$ & $^{+ 0.3}_{- 0.3}$ 
    & $^{+ 2.6}_{- 1.4}$  & $^{+ 2.5}_{- 0.2}$ 
 \\[2pt]
   500.0 &
    0.0901 & $^{+ 3.7}_{- 1.5}$ & $\pm 0.6$
    & $^{+ 1.2}_{- 1.2}$ & $^{+ 0.2}_{- 0.2}$ & $^{+ 0.3}_{- 0.3}$ 
    & $^{+ 1.9}_{- 0.3}$  & $^{+ 2.9}_{- 0.6}$ 
 \\[2pt]
   625.0 &
    0.0890 & $^{+ 3.9}_{- 1.8}$ & $\pm 0.5$
   & $^{+ 1.1}_{- 1.1}$ & $^{+ 0.1}_{- 0.1}$ & $^{+ 0.3}_{- 0.4}$ 
    & $^{+ 1.7}_{- 0.3}$  & $^{+ 3.3}_{- 1.3}$ 
 \\[2pt]
 800.0 &  0.0850 & $^{+ 5.9}_{- 2.2}$ & $\pm 0.6$
  & $^{+ 1.0}_{- 1.1}$ & $^{+ 0.1}_{- 0.1}$ & $^{+ 0.4}_{- 0.4}$ 
  & $^{+ 4.6}_{- 0.2}$ & $^{+ 3.5}_{- 1.8}$ 
 \\[2pt]
  1000 &
    0.0856 & $^{+ 4.0}_{- 2.7}$ & $\pm 1.2$
    & $^{+ 1.1}_{- 1.1}$ & $^{+ 0.1}_{- 0.1}$ & $^{+ 0.4}_{- 0.4}$
     & $^{+ 1.4}_{- 0.4}$  & $^{+ 3.4}_{- 2.0}$ 
 \\[2pt]
  1225 &  0.0790 & $^{+ 4.6}_{- 3.5}$ & $\pm 2.5$
  & $^{+ 1.2}_{- 1.2}$ & $^{+ 0.1}_{- 0.1}$ & $^{+ 0.5}_{- 0.5}$ 
    & $^{+ 1.6}_{- 0.4}$& $^{+ 3.2}_{- 1.9}$ 
 \\[2pt]
1675 &  0.0723 & $^{+ 7.0}_{- 8.6}$ & $\pm 6.1$
  & $^{+ 1.3}_{- 1.2}$ & $<\pm0.1$ & $^{+ 0.5}_{- 0.5}$ 
   & $^{+ 1.7}_{- 5.1}$& $^{+ 2.8}_{- 1.6}$ 
\\[2pt]
\hline \hline
\end{tabular}
\end{table*}

In addition, assuming the validity of the RGE, all 18 data points
in $0<y^*<0.5$ and $0.5<y^* < 1.0$ for $\Dphimax = 7\pi/8$
are used to extract a combined $\asmz$ result.
The combined fit (for MMHT2014 PDFs at the default scale) 
gives $\chi^2 = 21.7$ for 17 degrees of freedom
and a result of $\asmz = 0.1127$
(the uncertainties are detailed in Table~\ref{tab:asmz}).
The fit is then repeated for the 
CT14, NNPDFv2.3, ABMP16, and HERAPDF~2.0 PDF sets, 
for which the $\asmz$ results differ by 
$+0.0001$, $+0.0022$, $+0.0026$, and $+0.0029$, respectively.
Fits for various choices of $\mu_R$ and $\mu_F$ result
in variations of the $\asmz$ results between $-0.0019$ and $+0.0052$.

Further dependence of the $\as$ results on some of the analysis choices
is investigated in a series of systematic studies.

\begin{itemize}

\item{Changing the $\Dphimax$ requirement} \\
Based on the criteria outlined in Section~\ref{sec:criteria} it was decided
to use the data for $\Dphimax = 7\pi/8$ in the $\as$ analysis.
If, instead, the data with $\Dphimax = 5\pi/6$ are used,
the $\asmz$ result changes by $+0.0052$ to $\asmz = 0.1179$,
with an uncertainty of $+0.0065$ and $-0.0045$ due to the scale dependence.

\item{Extending the $y^*$ region} \\
For the central $\as$ results, the data points with
$1 < y^* < 2$ are excluded.
If $\asmz$ is determined only from the data points
for $1 < y^* < 2$ (with $\Dphimax = 7\pi/8$)
the $\asmz$ result changes by $-0.0018$,
with an increased scale dependence, 
to $\asmz = 0.1109^{+0.0071}_{-0.0031}$
with $\chi^2 = 13.8$ for seven degrees of freedom.
If the data points for $1 < y^* < 2$ are combined with 
those for $0 < y^* < 0.5$ and $0.5 < y^* < 1$,
the result is $\asmz = 0.1135^{+0.0051}_{-0.0025}$.

\item {Smoothing the systematic uncertainties} \\
In the experimental measurement, the systematic 
uncertainties that are correlated between different data points were 
smoothed in order to avoid double counting of statistical fluctuations.
For this purpose, the systematic uncertainties were fitted with a linear function in 
$\log{(\Ht/\mbox{\GeV})}$.
If, alternatively, a quadratic function is used, the central $\asmz$ 
result changes by $-0.0006$, and the experimental 
uncertainty is changed from $^{+0.0018}_{-0.0017}$
to $^{+0.0017}_{-0.0016}$.

\item{Stronger correlations of experimental uncertainties} \\
The largest experimental uncertainties are due to the jet energy calibration.
These are represented by contributions from 57 independent sources.
Some of the correlations are estimated on the basis of prior assumptions.
In a study of the systematic effects these assumptions are varied, resulting
in an alternative scenario with stronger correlations between 
some of these sources. 
This changes the combined $\asmz$ result by $-0.0004$,
while the experimental correlated uncertainty is reduced from
$^{+ 0.0018}_{- 0.0017}$ to  $^{+ 0.0012}_{- 0.0013}$.

\begin{table*}
\renewcommand{\arraystretch}{1.1}
\caption{\label{tab:asmz}
Fit result for $\asmz$, determined from the $\Rdphi$ data
for $\Dphimax = 7\pi/8$ with $0.0<\ystar<0.5$ and $0.5<\ystar<1.0$.
All uncertainties have been multiplied by a factor of $10^3$.}
\begin{tabular}{cccccccc}
\hline \hline
 $\asmz$  &
Total  &
Statistical & Experimental & Non-perturb. & MMHT2014 & PDF set & $\murf$ \\
 & uncert. & & correlated &
corrections & uncertainty & & variation \\[2pt]
\hline
  0.1127 & $^{+ 6.3}_{- 2.7}$ & $\pm 0.5$
   & $^{+ 1.8}_{- 1.7}$ & $^{+ 0.3}_{- 0.1}$ & $^{+ 0.6}_{- 0.6}$
 & $^{+ 2.9}_{- 0.0}$  & $^{+ 5.2}_{- 1.9}$ \\
[2pt]
\hline \hline
\end{tabular}
\end{table*}

\item {Treatment of non-perturbative corrections} \\
The central $\as$ results are obtained using
the average values of the non-perturbative corrections
from \pythia\ tunes ABT1 and DW, and the spread between the
average and the individual models is taken as a correlated uncertainty,
which is treated in the Hessian approach by fitting a corresponding
nuisance parameter.
Alternatively, the $\asmz$ result is also extracted by
fixing the values for the non-perturbative corrections
to the individual model predictions from \herwig\ (default)
and \pythia\ with tunes AMBT1, DW, S Global, and A, 
and to unity (corresponding to zero non-perturbative corrections).
The corresponding changes of the $\asmz$ result
for the different choices are between $-0.0004$ and $+0.0011$.

\item {Choice of $n_\textrm{f}=6$ versus $n_\textrm{f}=5$} \\
The choice of $n_\textrm{f}=6$ corresponds to the rather extreme 
approximation in which the top quark is included as a massless quark
in the pQCD calculation. 
The effect of using $n_\textrm{f} =6$ 
instead of $n_\textrm{f} =5$
in the pQCD matrix elements and the RGE 
and the corresponding impact on $\Rdphi$ 
are discussed in Appendix~\ref{sec:top}.
The effects on the extracted $\as$ results are also studied
and are found to be between 
$+1.3\%$ (at low \Ht) and $-1.1\%$
(at high \Ht) for the nine $\asq$ results.
The combined $\asmz$ result changes by $-0.0006$ from
$0.1127$ (for $n_\textrm{f}=5$) to $0.1121$ (for $n_\textrm{f}=6$).

\item {A scan of the renormalization scale dependence} \\
Unlike all other uncertainties which are treated in the Hessian approach, 
the uncertainty due to the renormalization and factorization scale dependence 
is obtained from individual fits in which both scales are set to fixed values.
To ensure that the largest variation may not occur at intermediate
values, a scan of the renormalization scale dependence 
in finer steps is made.
For each of the three variations of $\muf$ by factors of
$x_{\muf} = 0.5$, $1.0$, $2.0$,
the renormalization scale is varied by nine logarithmically equal-spaced 
factors of
$x_{\mur} =0.5$, $0.596$, $0.708$, $0.841$, $1.0$, $1.189$, $1.413$, $1.679$, 
and $2.0$.

It is seen that the largest upward variation (of $+0.0052$)
is obtained for the correlated variation $x_{\mur} = x_{\muf} = 2.0$.
The lowest variation (of $-0.0027$)
is obtained for the anti-correlated variation
$x_{\mur}=0.5$ and $x_{\muf} = 2.0$, which is, however,
outside the range $0.5 \le x_{\mur}/x_{\muf} \le 2$.
The lowest variation within this range
($-0.0014$) is obtained for 
$x_{\mur}=0.5$ and $x_{\muf} = 1.0$.

\item{Effects of the Hessian method} \\
In the Hessian approach, a fit can explore the multi-dimensional
uncertainty space to find the $\chi^2$ minimum at values of the 
nuisance parameters associated to the sources of systematic uncertainties,
that do not represent the best knowledge
of the corresponding sources.
While in this analysis the shifts of the nuisance parameters are 
all small, it is still interesting to study their effects on the 
$\as$ fit results.
Therefore, the $\asmz$ extraction is repeated, initially including 
the uncorrelated (i.e.\ statistical) uncertainties only.
Then, step by step, the experimental correlated uncertainties,
the uncertainties of the non-perturbative corrections, and the PDF uncertainties
are included.
These fits produce $\asmz$ results that differ by less than 
$\pm 0.0004$ from the central result.

\end{itemize}

These systematic studies show that the $\as$ results
are rather independent of the analysis choices and 
demonstrate the stability of the $\as$ extraction procedure.
These variations are not treated as additional uncertainties 
because their resulting effects are smaller than the other 
theoretical uncertainties.
The largest variation of the $\asmz$ result, by 
$+0.0052$, is obtained 
when using the data with $\Dphimax = 5\pi/6$ instead of $\Dphimax=7\pi/8$.
This difference may be due to different higher-order corrections
to the NLO pQCD results for different $\Dphimax$ values.
This assumption is consistent with the observed scale dependence of
the $\asmz$ results, within which the results for both choices 
of $\Dphimax$ agree ($0.1127+0.0052$ versus $0.1179-0.0045$
for $\Dphimax = 5\pi/6$ and $7\pi/8$, respectively).
It is therefore concluded from the systematic studies that no further 
uncertainties need to be assigned.

\begin{figure}
 \centering
\includegraphics[scale=1.15]{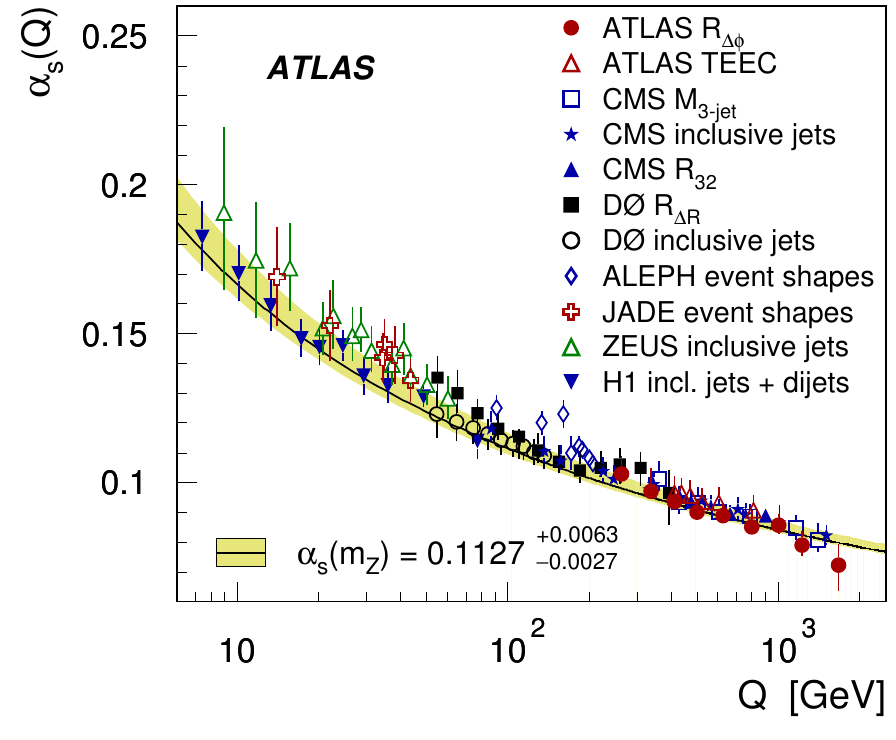}
\caption{\label{fig:alphas2} The $\asq$ results from this analysis
   in the range of $262 < Q < 1675\,$\GeV, compared 
   to the results of previous $\as$ determinations 
   from jet data in other experiments 
   at $5<Q<1508\,$\GeV~\cite{Andreev:2017vxu,Abramowicz:2012jz,Bethke:2008hf,Dissertori:2009ik,Abazov:2009nc,Abazov:2012lua,Chatrchyan:2013txa,Khachatryan:2014waa,Khachatryan:2016mlc,CMS:2014mna,Aaboud:2017fml}.
   Also shown is the prediction of the RGE for the 
   $\asmz$ result obtained 
   from the $\Rdphi$ data in this analysis.
}
\end{figure}

The final result from the combined fit is $\asmz = 0.1127^{+0.0063}_{-0.0027}$
with the individual uncertainty contributions given in Table~\ref{tab:asmz}.
This result and the corresponding RGE prediction
are also shown in Figure~\ref{fig:alphas1}.
For all $\as$ results in Tables~\ref{tab:asq} and~\ref{tab:asmz}, 
the uncertainties are dominated by the $\mur$ dependence 
of the NLO pQCD calculation.

Within the uncertainties, the $\asmz$ result is consistent 
with the current world average value of 
$\asmz = 0.1181 \pm 0.0011$~\cite{Olive:2016xmw}
and with recent $\as$ results from multi-jet cross-section ratio 
measurements in hadron collisions, 
namely from the D\O\ measurement of $\Rdr$~\cite{Abazov:2012lua} 
($\asmz=0.1191^{+0.0048}_{-0.0071}$),
and from the CMS measurements of $\Rtt$~\cite{Chatrchyan:2013txa}
($\asmz=0.1148\pm0.0055$),
the inclusive jet cross section~\cite{Khachatryan:2014waa,Khachatryan:2016mlc}
($\asmz=0.1185^{+0.0063}_{-0.0042}$, $\asmz=0.1164^{+0.0060}_{-0.0043}$),
and the three-jet cross section~\cite{CMS:2014mna}
($\asmz=0.1171^{+0.0074}_{-0.0049}$),
and the ATLAS measurement of transverse energy--energy 
correlations~\cite{Aaboud:2017fml}
($\asmz=0.1162^{+0.0085}_{-0.0071}$),
with comparable uncertainties.
The compatibility of the results of this analysis, based on the
measurements of $\Rdphi$, with the world average
value of $\asmz$ is demonstrated in Appendix~\ref{sec:asworld}.

The individual $\asq$ results are compared in Figure~\ref{fig:alphas2}
with previously published $\as$ results obtained from jet measurements~\cite{Dissertori:2009ik,Abazov:2009nc,Abazov:2012lua,Bethke:2008hf,Andreev:2017vxu,Abramowicz:2012jz,Chatrchyan:2013txa,Khachatryan:2014waa,Khachatryan:2016mlc,CMS:2014mna,Aaboud:2017fml}
and with the RGE prediction for the combined $\asmz$ result 
obtained in this analysis.
The new results agree with previous $\asq$ results 
in the region of overlap, and extend the pQCD tests
to momentum transfers up to 1.6~\TeV,
where RGE predictions are consistent with the $\asq$ results,
as discussed in Appendix~\ref{sec:asslope}.

\section{Summary \label{sec:summary}}

The multi-jet cross-section ratio $\Rdphi$ 
is measured at the LHC.
The quantity $\Rdphi$ specifies the fraction of the inclusive dijet
events in which the azimuthal opening angle of the two jets with the highest
transverse momenta is less than a given value of the parameter $\Dphimax$.
The $\Rdphi$ results, measured in 20.2 fb$^{-1}$ of
$pp$ collisions at $\sqrt{s}=8\,$\TeV\/ with the ATLAS detector,
are presented as a function of three variables:
the total transverse momentum \Ht, 
the dijet rapidity interval $y^*$, and the parameter $\Dphimax$.
The \Ht and $y^*$ dependences of the data are well-described 
by theoretical predictions 
based on NLO pQCD (for $\Dphimax = 7\pi/8$, $5\pi/6$, and $3\pi/4$),
or LO pQCD (for $\Dphimax = 2\pi/3$),
with corrections for non-perturbative effects.
Based on the data points for  $\Dphimax = 7\pi/8$ with
$0< y^* < 0.5$ and  $0.5< y^* < 1$, nine $\as$ results are determined,
at a scale of $Q = \Ht/2$, over the range of $262< Q < 1675\,$\GeV.
The $\asq$ results are consistent with the predictions
of the RGE, and a combined analysis results
in a value of $\asmz = 0.1127^{+0.0063}_{-0.0027}$,
where the uncertainty is dominated by the scale dependence
of the NLO pQCD predictions.

\section*{Acknowledgments}


We thank CERN for the very successful operation of the LHC, as well as the
support staff from our institutions without whom ATLAS could not be
operated efficiently.

We acknowledge the support of ANPCyT, Argentina; YerPhI, Armenia; ARC, Australia; BMWFW and FWF, Austria; ANAS, Azerbaijan; SSTC, Belarus; CNPq and FAPESP, Brazil; NSERC, NRC and CFI, Canada; CERN; CONICYT, Chile; CAS, MOST and NSFC, China; COLCIENCIAS, Colombia; MSMT CR, MPO CR and VSC CR, Czech Republic; DNRF and DNSRC, Denmark; IN2P3-CNRS, CEA-DRF/IRFU, France; SRNSFG, Georgia; BMBF, HGF, and MPG, Germany; GSRT, Greece; RGC, Hong Kong SAR, China; ISF, I-CORE and Benoziyo Center, Israel; INFN, Italy; MEXT and JSPS, Japan; CNRST, Morocco; NWO, Netherlands; RCN, Norway; MNiSW and NCN, Poland; FCT, Portugal; MNE/IFA, Romania; MES of Russia and NRC KI, Russian Federation; JINR; MESTD, Serbia; MSSR, Slovakia; ARRS and MIZ\v{S}, Slovenia; DST/NRF, South Africa; MINECO, Spain; SRC and Wallenberg Foundation, Sweden; SERI, SNSF and Cantons of Bern and Geneva, Switzerland; MOST, Taiwan; TAEK, Turkey; STFC, United Kingdom; DOE and NSF, United States of America. In addition, individual groups and members have received support from BCKDF, the Canada Council, CANARIE, CRC, Compute Canada, FQRNT, and the Ontario Innovation Trust, Canada; EPLANET, ERC, ERDF, FP7, Horizon 2020 and Marie Sk{\l}odowska-Curie Actions, European Union; Investissements d'Avenir Labex and Idex, ANR, R{\'e}gion Auvergne and Fondation Partager le Savoir, France; DFG and AvH Foundation, Germany; Herakleitos, Thales and Aristeia programmes co-financed by EU-ESF and the Greek NSRF; BSF, GIF and Minerva, Israel; BRF, Norway; CERCA Programme Generalitat de Catalunya, Generalitat Valenciana, Spain; the Royal Society and Leverhulme Trust, United Kingdom.

The crucial computing support from all WLCG partners is acknowledged gratefully, in particular from CERN, the ATLAS Tier-1 facilities at TRIUMF (Canada), NDGF (Denmark, Norway, Sweden), CC-IN2P3 (France), KIT/GridKA (Germany), INFN-CNAF (Italy), NL-T1 (Netherlands), PIC (Spain), ASGC (Taiwan), RAL (UK) and BNL (USA), the Tier-2 facilities worldwide and large non-WLCG resource providers. Major contributors of computing resources are listed in Ref.~\cite{ATL-GEN-PUB-2016-002}.

\clearpage
\appendix
\part*{Appendix}
\addcontentsline{toc}{part}{Appendix}


\section{Effects of top quark contributions 
         on the pQCD predictions \label{sec:top}}

There are two ways in which contributions from top quarks affect
the pQCD predictions for $\Rdphi$.
Firstly, 
the pQCD predictions based on matrix elements for massless quarks
also depend on the number of quark flavors 
in gluon splitting ($g \rightarrow q\bar{q}$), $n_\textrm{f}$,
which affects the tree-level matrix elements 
and their real and virtual corrections,
as well as the RGE predictions.
The pQCD predictions for the central analysis are obtained
for $n_\textrm{f}=5$.
The effects for the measured quantity $\Rdphi$ for the choice $n_\textrm{f} =6$ 
are computed in this appendix.
Secondly,
since the decay products of hadronically decaying (anti-)top quarks 
are sometimes reconstructed as multiple jets, 
the $\ord(\ass)$ $t\bar{t}$ production process also contributes 
to three-jet topologies.
Since this contribution is of lower order in $\as$
as compared to the pQCD $\ord(\asss)$ three-jet production processes,
it is a "super-leading" contribution, which is formally more important.
This potentially large contribution and the corresponding effects
for $\Rdphi$ are also estimated in this appendix.

In a pQCD calculation in which quark masses are properly taken into account,
the contributions from the massive top quark arise naturally
at higher momentum transfers, according to the available phase space.
In calculations based on matrix elements for massless quarks,
$n_\textrm{f}$ is a parameter in the calculation.
For jet production at the LHC, the alternatives are $n_\textrm{f} =5$,
i.e.\ ignoring the contributions from $g \rightarrow t\bar{t}$ processes
(which is the central choice for this analysis),
or $n_\textrm{f} =6$, i.e.\ treating the top quark as a sixth massless quark.
The relative difference between the two alternatives is evaluated
from the effects due to the RGE and the matrix elements.
For this purpose, the 2-loop solution of the RGE for $n_\textrm{f}=5$ is replaced
by the 2-loop solutions for $n_\textrm{f}=5$ and $n_\textrm{f}=6$ with 
1-loop matching~\cite{Chetyrkin:1997sg}
at the pole mass of the top quark $m_\mathrm{top}^\mathrm{pole}$,
assuming that  $m_\mathrm{top}^\mathrm{pole}$ is equal to the 
world average of the measured ``Monte Carlo mass'' of 
$173.21\,$\GeV~\cite{Olive:2016xmw}.
In addition, the matrix elements are recomputed for $n_\textrm{f}=6$.
For a fixed value of $\asmz = 0.118$, the corresponding effects 
for the pQCD predictions for $\Rdphi$ are in the range
of $-1\%$ to $+2\%$.

The effects on $\Rdphi$ due to the contributions from hadronic decays
of $t\bar{t}$ final states are estimated using 
\powheg~\cite{Nason:2004rx} (for the pQCD matrix elements)
interfaced with \pythia\ (for the parton shower, underlying event, 
and hadronization) 
and CTEQ6L1 PDFs~\cite{Pumplin:2002vw}.
It is seen that the $t\bar{t}$ process contributes
0.003--0.2\% to the denominator of $\Rdphi$
(the inclusive dijet cross section),
and 0.006--0.5\% to the numerator (with $\Dphimax = 7\pi/8$).
The effects for the ratio $\Rdphi$ are 0--0.5\%
in the analysis phase space,  
and there are no systematic trends in the considered distributions
within the statistical uncertainties
of the generated \powheg\ event sample.
Since this effect is about four to eight times smaller 
than the typical uncertainty due to the renormalization scale dependence,
the corresponding effects on $\as$ are not investigated further.

\section{Data tables \label{sec:datatable}}

The results of the $\Rdphi$ measurements are listed 
in Tables~\ref{tab:data1}--\ref{tab:data4},
together with their relative statistical and systematic 
uncertainties.
A detailed list of the individual contributions from all sources of 
correlated uncertainties is provided in Ref.~\cite{suppl}.

\begin{table}[t]
\renewcommand{\arraystretch}{1.1}
\centering
\caption{\label{tab:data1}
The $\Rdphi$ measurement results
for $\Dphimax=7\pi/8$
with their relative statistical and systematic uncertainties.
}
\begin{tabular}{r r r c r r}
\hline \hline
 \multicolumn{1}{c}{\Ht}  &  \multicolumn{1}{c}{$y^*$} & \multicolumn{1}{c}{$\Rdphi$} & \multicolumn{1}{c}{Stat.~uncert.} &
 \multicolumn{2}{c}{Syst.~uncert.} \\
 \multicolumn{1}{c}{[\GeV]} &   & & \multicolumn{1}{c}{[\%]} & \multicolumn{2}{c}{[\%]} \\
\hline
 $ 450$--$ 600$  &  $ 0.0$--$ 0.5$  &  $ 1.88 \cdot 10^{-1}$  &  $\pm  2.2$  &  $+  1.8$  &  $-  1.7$ \\
 $ 600$--$ 750$  &  $ 0.0$--$ 0.5$  &  $ 1.85 \cdot 10^{-1}$  &  $\pm  2.2$  &  $+  1.6$  &  $-  1.5$ \\
 $ 750$--$ 900$  &  $ 0.0$--$ 0.5$  &  $ 1.82 \cdot 10^{-1}$  &  $\pm  1.3$  &  $+  1.4$  &  $-  1.4$ \\
 $ 900$--$1100$  &  $ 0.0$--$ 0.5$  &  $ 1.67 \cdot 10^{-1}$  &  $\pm  0.9$  &  $+  1.3$  &  $-  1.3$ \\
 $1100$--$1400$  &  $ 0.0$--$ 0.5$  &  $ 1.56 \cdot 10^{-1}$  &  $\pm  0.7$  &  $+  1.2$  &  $-  1.2$ \\
 $1400$--$1800$  &  $ 0.0$--$ 0.5$  &  $ 1.36 \cdot 10^{-1}$  &  $\pm  1.0$  &  $+  1.2$  &  $-  1.2$ \\
 $1800$--$2200$  &  $ 0.0$--$ 0.5$  &  $ 1.25 \cdot 10^{-1}$  &  $\pm  1.9$  &  $+  1.2$  &  $-  1.3$ \\
 $2200$--$2700$  &  $ 0.0$--$ 0.5$  &  $ 1.02 \cdot 10^{-1}$  &  $\pm  4.1$  &  $+  1.3$  &  $-  1.4$ \\
 $2700$--$4000$  &  $ 0.0$--$ 0.5$  &  $ 0.82 \cdot 10^{-1}$  &  $\pm  9.9$  &  $+  1.5$  &  $-  1.7$ \\
\hline
 $ 450$--$ 600$  &  $ 0.5$--$ 1.0$  &  $ 1.97 \cdot 10^{-1}$  &  $\pm  2.2$  &  $+  1.5$  &  $-  1.6$ \\
 $ 600$--$ 750$  &  $ 0.5$--$ 1.0$  &  $ 2.04 \cdot 10^{-1}$  &  $\pm  2.3$  &  $+  1.3$  &  $-  1.4$ \\
 $ 750$--$ 900$  &  $ 0.5$--$ 1.0$  &  $ 1.94 \cdot 10^{-1}$  &  $\pm  1.3$  &  $+  1.2$  &  $-  1.3$ \\
 $ 900$--$1100$  &  $ 0.5$--$ 1.0$  &  $ 1.83 \cdot 10^{-1}$  &  $\pm  0.8$  &  $+  1.2$  &  $-  1.2$ \\
 $1100$--$1400$  &  $ 0.5$--$ 1.0$  &  $ 1.73 \cdot 10^{-1}$  &  $\pm  0.8$  &  $+  1.3$  &  $-  1.2$ \\
 $1400$--$1800$  &  $ 0.5$--$ 1.0$  &  $ 1.59 \cdot 10^{-1}$  &  $\pm  1.1$  &  $+  1.4$  &  $-  1.3$ \\
 $1800$--$2200$  &  $ 0.5$--$ 1.0$  &  $ 1.44 \cdot 10^{-1}$  &  $\pm  2.3$  &  $+  1.7$  &  $-  1.5$ \\
 $2200$--$2700$  &  $ 0.5$--$ 1.0$  &  $ 1.28 \cdot 10^{-1}$  &  $\pm  5.4$  &  $+  1.9$  &  $-  1.7$ \\
 $2700$--$4000$  &  $ 0.5$--$ 1.0$  &  $ 1.13 \cdot 10^{-1}$  &  $\pm 16$  &  $+  2.4$  &  $-  2.0$ \\
\hline
 $ 450$--$ 600$  &  $ 1.0$--$ 2.0$  &  $ 2.42 \cdot 10^{-1}$  &  $\pm  2.3$  &  $+  2.3$  &  $-  1.0$ \\
 $ 600$--$ 750$  &  $ 1.0$--$ 2.0$  &  $ 2.40 \cdot 10^{-1}$  &  $\pm  2.5$  &  $+  1.9$  &  $-  1.1$ \\
 $ 750$--$ 900$  &  $ 1.0$--$ 2.0$  &  $ 2.54 \cdot 10^{-1}$  &  $\pm  1.5$  &  $+  1.7$  &  $-  1.2$ \\
 $ 900$--$1100$  &  $ 1.0$--$ 2.0$  &  $ 2.40 \cdot 10^{-1}$  &  $\pm  1.1$  &  $+  1.6$  &  $-  1.4$ \\
 $1100$--$1400$  &  $ 1.0$--$ 2.0$  &  $ 2.33 \cdot 10^{-1}$  &  $\pm  1.0$  &  $+  1.6$  &  $-  1.7$ \\
 $1400$--$1800$  &  $ 1.0$--$ 2.0$  &  $ 2.18 \cdot 10^{-1}$  &  $\pm  1.8$  &  $+  1.6$  &  $-  2.2$ \\
 $1800$--$2200$  &  $ 1.0$--$ 2.0$  &  $ 2.22 \cdot 10^{-1}$  &  $\pm  4.4$  &  $+  1.6$  &  $-  2.7$ \\
 $2200$--$2700$  &  $ 1.0$--$ 2.0$  &  $ 1.96 \cdot 10^{-1}$  &  $\pm 14$  &  $+  1.7$  &  $-  3.1$ \\
\hline \hline
\end{tabular}
\end{table}

\begin{table}[t]
\renewcommand{\arraystretch}{1.1}
\centering
\caption{\label{tab:data2}
The $\Rdphi$ measurement results 
for $\Dphimax=5\pi/6$
with their relative statistical and systematic uncertainties.
}
\begin{tabular}{r r r c r r}
\hline \hline
 \multicolumn{1}{c}{\Ht}  &  \multicolumn{1}{c}{$y^*$} & \multicolumn{1}{c}{$\Rdphi$} & \multicolumn{1}{c}{Stat.~uncert.} &
 \multicolumn{2}{c}{Syst.~uncert.} \\
 \multicolumn{1}{c}{[\GeV]} &   & & \multicolumn{1}{c}{[\%]} & \multicolumn{2}{c}{[\%]} \\
\hline
 $ 450$--$ 600$  &  $ 0.0$--$ 0.5$  &  $ 1.22 \cdot 10^{-1}$  &  $\pm  2.8$  &  $+  2.0$  &  $-  1.9$ \\
 $ 600$--$ 750$  &  $ 0.0$--$ 0.5$  &  $ 1.13 \cdot 10^{-1}$  &  $\pm  2.9$  &  $+  1.7$  &  $-  1.7$ \\
 $ 750$--$ 900$  &  $ 0.0$--$ 0.5$  &  $ 1.10 \cdot 10^{-1}$  &  $\pm  1.7$  &  $+  1.5$  &  $-  1.6$ \\
 $ 900$--$1100$  &  $ 0.0$--$ 0.5$  &  $ 1.00 \cdot 10^{-1}$  &  $\pm  1.3$  &  $+  1.4$  &  $-  1.5$ \\
 $1100$--$1400$  &  $ 0.0$--$ 0.5$  &  $ 0.92 \cdot 10^{-1}$  &  $\pm  1.0$  &  $+  1.2$  &  $-  1.5$ \\
 $1400$--$1800$  &  $ 0.0$--$ 0.5$  &  $ 0.78 \cdot 10^{-1}$  &  $\pm  1.4$  &  $+  1.2$  &  $-  1.5$ \\
 $1800$--$2200$  &  $ 0.0$--$ 0.5$  &  $ 0.72 \cdot 10^{-1}$  &  $\pm  2.6$  &  $+  1.2$  &  $-  1.7$ \\
 $2200$--$2700$  &  $ 0.0$--$ 0.5$  &  $ 0.55 \cdot 10^{-1}$  &  $\pm  5.7$  &  $+  1.3$  &  $-  1.9$ \\
 $2700$--$4000$  &  $ 0.0$--$ 0.5$  &  $ 0.51 \cdot 10^{-1}$  &  $\pm 13$  &  $+  1.6$  &  $-  2.3$ \\
\hline
 $ 450$--$ 600$  &  $ 0.5$--$ 1.0$  &  $ 1.33 \cdot 10^{-1}$  &  $\pm  2.9$  &  $+  1.5$  &  $-  1.8$ \\
 $ 600$--$ 750$  &  $ 0.5$--$ 1.0$  &  $ 1.27 \cdot 10^{-1}$  &  $\pm  3.1$  &  $+  1.4$  &  $-  1.5$ \\
 $ 750$--$ 900$  &  $ 0.5$--$ 1.0$  &  $ 1.18 \cdot 10^{-1}$  &  $\pm  1.8$  &  $+  1.3$  &  $-  1.3$ \\
 $ 900$--$1100$  &  $ 0.5$--$ 1.0$  &  $ 1.11 \cdot 10^{-1}$  &  $\pm  1.2$  &  $+  1.3$  &  $-  1.2$ \\
 $1100$--$1400$  &  $ 0.5$--$ 1.0$  &  $ 1.03 \cdot 10^{-1}$  &  $\pm  1.2$  &  $+  1.4$  &  $-  1.2$ \\
 $1400$--$1800$  &  $ 0.5$--$ 1.0$  &  $ 0.93 \cdot 10^{-1}$  &  $\pm  1.5$  &  $+  1.6$  &  $-  1.3$ \\
 $1800$--$2200$  &  $ 0.5$--$ 1.0$  &  $ 0.85 \cdot 10^{-1}$  &  $\pm  3.2$  &  $+  1.9$  &  $-  1.4$ \\
 $2200$--$2700$  &  $ 0.5$--$ 1.0$  &  $ 0.74 \cdot 10^{-1}$  &  $\pm  7.3$  &  $+  2.2$  &  $-  1.6$ \\
\hline
 $ 450$--$ 600$  &  $ 1.0$--$ 2.0$  &  $ 1.58 \cdot 10^{-1}$  &  $\pm  2.9$  &  $+  3.1$  &  $-  1.0$ \\
 $ 600$--$ 750$  &  $ 1.0$--$ 2.0$  &  $ 1.54 \cdot 10^{-1}$  &  $\pm  3.3$  &  $+  2.5$  &  $-  0.9$ \\
 $ 750$--$ 900$  &  $ 1.0$--$ 2.0$  &  $ 1.62 \cdot 10^{-1}$  &  $\pm  2.3$  &  $+  2.1$  &  $-  1.1$ \\
 $ 900$--$1100$  &  $ 1.0$--$ 2.0$  &  $ 1.53 \cdot 10^{-1}$  &  $\pm  1.6$  &  $+  1.9$  &  $-  1.5$ \\
 $1100$--$1400$  &  $ 1.0$--$ 2.0$  &  $ 1.47 \cdot 10^{-1}$  &  $\pm  1.4$  &  $+  1.8$  &  $-  2.2$ \\
 $1400$--$1800$  &  $ 1.0$--$ 2.0$  &  $ 1.36 \cdot 10^{-1}$  &  $\pm  2.6$  &  $+  1.8$  &  $-  3.1$ \\
 $1800$--$2200$  &  $ 1.0$--$ 2.0$  &  $ 1.41 \cdot 10^{-1}$  &  $\pm  5.8$  &  $+  1.9$  &  $-  3.9$ \\
 $2200$--$2700$  &  $ 1.0$--$ 2.0$  &  $ 1.35 \cdot 10^{-1}$  &  $\pm 18$  &  $+  2.0$  &  $-  4.7$ \\
\hline \hline
\end{tabular}
\end{table}

\begin{table}[t]
\renewcommand{\arraystretch}{1.1}
\centering
\caption{\label{tab:data3}
The $\Rdphi$ measurement results 
for $\Dphimax=3\pi/4$
with their relative statistical and systematic uncertainties.
}
\begin{tabular}{r r r c r r}
\hline \hline
 \multicolumn{1}{c}{\Ht}  &  \multicolumn{1}{c}{$y^*$} & \multicolumn{1}{c}{$\Rdphi$} & \multicolumn{1}{c}{Stat.~uncert.} &
 \multicolumn{2}{c}{Syst.~uncert.} \\
 \multicolumn{1}{c}{[\GeV]} &   & & \multicolumn{1}{c}{[\%]} & \multicolumn{2}{c}{[\%]} \\
\hline
 $ 450$--$ 600$  &  $ 0.0$--$ 0.5$  &  $ 4.35 \cdot 10^{-2}$  &  $\pm  5.0$  &  $+  3.4$  &  $-  2.4$ \\
 $ 600$--$ 750$  &  $ 0.0$--$ 0.5$  &  $ 3.67 \cdot 10^{-2}$  &  $\pm  5.9$  &  $+  3.0$  &  $-  2.1$ \\
 $ 750$--$ 900$  &  $ 0.0$--$ 0.5$  &  $ 3.55 \cdot 10^{-2}$  &  $\pm  4.6$  &  $+  2.6$  &  $-  1.9$ \\
 $ 900$--$1100$  &  $ 0.0$--$ 0.5$  &  $ 3.24 \cdot 10^{-2}$  &  $\pm  3.9$  &  $+  2.3$  &  $-  1.8$ \\
 $1100$--$1400$  &  $ 0.0$--$ 0.5$  &  $ 2.84 \cdot 10^{-2}$  &  $\pm  2.5$  &  $+  2.0$  &  $-  1.8$ \\
 $1400$--$1800$  &  $ 0.0$--$ 0.5$  &  $ 2.27 \cdot 10^{-2}$  &  $\pm  3.2$  &  $+  1.8$  &  $-  2.0$ \\
 $1800$--$2200$  &  $ 0.0$--$ 0.5$  &  $ 1.89 \cdot 10^{-2}$  &  $\pm  5.5$  &  $+  1.8$  &  $-  2.2$ \\
 $2200$--$2700$  &  $ 0.0$--$ 0.5$  &  $ 1.43 \cdot 10^{-2}$  &  $\pm 12$  &  $+  1.9$  &  $-  2.5$ \\
\hline
 $ 450$--$ 600$  &  $ 0.5$--$ 1.0$  &  $ 4.68 \cdot 10^{-2}$  &  $\pm  5.5$  &  $+  2.2$  &  $-  2.6$ \\
 $ 600$--$ 750$  &  $ 0.5$--$ 1.0$  &  $ 4.01 \cdot 10^{-2}$  &  $\pm  6.1$  &  $+  1.8$  &  $-  1.9$ \\
 $ 750$--$ 900$  &  $ 0.5$--$ 1.0$  &  $ 3.92 \cdot 10^{-2}$  &  $\pm  4.1$  &  $+  1.6$  &  $-  1.6$ \\
 $ 900$--$1100$  &  $ 0.5$--$ 1.0$  &  $ 3.61 \cdot 10^{-2}$  &  $\pm  2.9$  &  $+  1.5$  &  $-  1.4$ \\
 $1100$--$1400$  &  $ 0.5$--$ 1.0$  &  $ 3.31 \cdot 10^{-2}$  &  $\pm  3.3$  &  $+  1.6$  &  $-  1.3$ \\
 $1400$--$1800$  &  $ 0.5$--$ 1.0$  &  $ 2.90 \cdot 10^{-2}$  &  $\pm  3.4$  &  $+  2.1$  &  $-  1.3$ \\
 $1800$--$2200$  &  $ 0.5$--$ 1.0$  &  $ 2.44 \cdot 10^{-2}$  &  $\pm  6.7$  &  $+  2.5$  &  $-  1.5$ \\
 $2200$--$2700$  &  $ 0.5$--$ 1.0$  &  $ 2.17 \cdot 10^{-2}$  &  $\pm 14$  &  $+  3.0$  &  $-  1.8$ \\
\hline
 $ 450$--$ 600$  &  $ 1.0$--$ 2.0$  &  $ 6.02 \cdot 10^{-2}$  &  $\pm  5.1$  &  $+  5.8$  &  $-  2.5$ \\
 $ 600$--$ 750$  &  $ 1.0$--$ 2.0$  &  $ 5.68 \cdot 10^{-2}$  &  $\pm  5.7$  &  $+  4.8$  &  $-  2.4$ \\
 $ 750$--$ 900$  &  $ 1.0$--$ 2.0$  &  $ 5.71 \cdot 10^{-2}$  &  $\pm  4.6$  &  $+  4.1$  &  $-  2.7$ \\
 $ 900$--$1100$  &  $ 1.0$--$ 2.0$  &  $ 5.19 \cdot 10^{-2}$  &  $\pm  3.4$  &  $+  3.7$  &  $-  3.2$ \\
 $1100$--$1400$  &  $ 1.0$--$ 2.0$  &  $ 4.95 \cdot 10^{-2}$  &  $\pm  2.7$  &  $+  3.5$  &  $-  4.0$ \\
 $1400$--$1800$  &  $ 1.0$--$ 2.0$  &  $ 4.56 \cdot 10^{-2}$  &  $\pm  5.0$  &  $+  3.7$  &  $-  5.0$ \\
 $1800$--$2200$  &  $ 1.0$--$ 2.0$  &  $ 5.25 \cdot 10^{-2}$  &  $\pm 11$  &  $+  4.1$  &  $-  6.1$ \\
\hline \hline
\end{tabular}
\end{table}

\begin{table}[t]
\renewcommand{\arraystretch}{1.1}
\centering
\caption{\label{tab:data4}
The $\Rdphi$ measurement results 
for $\Dphimax=2\pi/3$
with their relative statistical and systematic uncertainties.
}
\begin{tabular}{r r r c r r}
\hline \hline
 \multicolumn{1}{c}{\Ht}  &  \multicolumn{1}{c}{$y^*$} & \multicolumn{1}{c}{$\Rdphi$} & \multicolumn{1}{c}{Stat.~uncert.} &
 \multicolumn{2}{c}{Syst.~uncert.} \\
 \multicolumn{1}{c}{[\GeV]} &   & & \multicolumn{1}{c}{[\%]} & \multicolumn{2}{c}{[\%]} \\
\hline
 $ 450$--$ 600$  &  $ 0.0$--$ 0.5$  &  $ 1.37 \cdot 10^{-2}$  &  $\pm  9.5$  &  $+  6.3$  &  $-  4.1$ \\
 $ 600$--$ 750$  &  $ 0.0$--$ 0.5$  &  $ 1.05 \cdot 10^{-2}$  &  $\pm 11$  &  $+  5.4$  &  $-  3.6$ \\
 $ 750$--$ 900$  &  $ 0.0$--$ 0.5$  &  $ 1.02 \cdot 10^{-2}$  &  $\pm 12$  &  $+  4.7$  &  $-  3.3$ \\
 $ 900$--$1100$  &  $ 0.0$--$ 0.5$  &  $ 0.87 \cdot 10^{-2}$  &  $\pm  8.9$  &  $+  4.1$  &  $-  3.2$ \\
 $1100$--$1400$  &  $ 0.0$--$ 0.5$  &  $ 0.70 \cdot 10^{-2}$  &  $\pm  6.0$  &  $+  3.5$  &  $-  3.2$ \\
 $1400$--$1800$  &  $ 0.0$--$ 0.5$  &  $ 0.48 \cdot 10^{-2}$  &  $\pm  7.8$  &  $+  3.2$  &  $-  3.3$ \\
 $1800$--$2200$  &  $ 0.0$--$ 0.5$  &  $ 0.38 \cdot 10^{-2}$  &  $\pm 13$  &  $+  3.2$  &  $-  3.7$ \\
\hline
 $ 450$--$ 600$  &  $ 0.5$--$ 1.0$  &  $ 1.45 \cdot 10^{-2}$  &  $\pm 11$  &  $+  3.9$  &  $-  4.4$ \\
 $ 600$--$ 750$  &  $ 0.5$--$ 1.0$  &  $ 1.07 \cdot 10^{-2}$  &  $\pm 12$  &  $+  2.7$  &  $-  2.5$ \\
 $ 750$--$ 900$  &  $ 0.5$--$ 1.0$  &  $ 1.14 \cdot 10^{-2}$  &  $\pm 11$  &  $+  2.1$  &  $-  1.8$ \\
 $ 900$--$1100$  &  $ 0.5$--$ 1.0$  &  $ 0.86 \cdot 10^{-2}$  &  $\pm  6.8$  &  $+  2.2$  &  $-  1.8$ \\
 $1100$--$1400$  &  $ 0.5$--$ 1.0$  &  $ 0.77 \cdot 10^{-2}$  &  $\pm  7.1$  &  $+  2.8$  &  $-  2.3$ \\
 $1400$--$1800$  &  $ 0.5$--$ 1.0$  &  $ 0.70 \cdot 10^{-2}$  &  $\pm  8.6$  &  $+  3.8$  &  $-  3.2$ \\
 $1800$--$2200$  &  $ 0.5$--$ 1.0$  &  $ 0.63 \cdot 10^{-2}$  &  $\pm 16$  &  $+  4.8$  &  $-  4.2$ \\
\hline
 $ 450$--$ 600$  &  $ 1.0$--$ 2.0$  &  $ 1.49 \cdot 10^{-2}$  &  $\pm 10$  &  $+  9.0$  &  $-  5.1$ \\
 $ 600$--$ 750$  &  $ 1.0$--$ 2.0$  &  $ 1.70 \cdot 10^{-2}$  &  $\pm 11$  &  $+  7.4$  &  $-  3.8$ \\
 $ 750$--$ 900$  &  $ 1.0$--$ 2.0$  &  $ 1.53 \cdot 10^{-2}$  &  $\pm  8.9$  &  $+  6.5$  &  $-  3.7$ \\
 $ 900$--$1100$  &  $ 1.0$--$ 2.0$  &  $ 1.29 \cdot 10^{-2}$  &  $\pm  7.5$  &  $+  6.2$  &  $-  4.3$ \\
 $1100$--$1400$  &  $ 1.0$--$ 2.0$  &  $ 1.12 \cdot 10^{-2}$  &  $\pm  6.6$  &  $+  6.6$  &  $-  5.9$ \\
 $1400$--$1800$  &  $ 1.0$--$ 2.0$  &  $ 1.02 \cdot 10^{-2}$  &  $\pm 12$  &  $+  7.6$  &  $-  8.0$ \\
 $1800$--$2200$  &  $ 1.0$--$ 2.0$  &  $ 1.61 \cdot 10^{-2}$  &  $\pm 20$  &  $+  8.8$  &  $- 10$ \\
\hline \hline
\end{tabular}
\end{table}

\clearpage
\section{Definition of $\chi^2$ \label{sec:chi2}}

Given is a set of experimental measurement results in bins $i$ 
of a given quantity
with central measurement results $d_i$ with 
statistical and uncorrelated systematic uncertainties 
$\sigma_{i,\mathrm{stat}}$ and
$\sigma_{i,\mathrm{uncorr}}$, respectively.
The experimental measurements are affected by 
various sources of correlated uncertainties,
and $\delta_{ij}(\epsilon_j)$ specifies the uncertainty 
of measurement $i$ due to the source $j$,
where $\epsilon_j$ is a Gaussian distributed random variable with
zero expectation value and unit width.
The $\delta_{ij}(\epsilon_j)$ specify the dependence of the measured 
result $i$ on the variation of the correlated uncertainty source
$j$ by $\epsilon_j$ standard deviations,
where $\epsilon_j = 0$ corresponds to the central value of the measurement
(i.e.\ $\delta_{ij}(\epsilon_j = 0)=0$),
while the relative uncertainties corresponding to plus/minus 
one standard deviation
are given by $\delta_{ij}(\epsilon_j=\pm1)=\Delta d_{ij}^\pm$.
From the central measurement result and the relative uncertainties 
$\Delta d_{ij}^\pm$, the continuous $\epsilon_j$ dependence of $\delta_{ij}(\epsilon_j)$
can be obtained using quadratic interpolation
\begin{equation*}
 \delta_{ij}(\epsilon_j) =  \epsilon_j 
\frac{\Delta d^+_{ij}-\Delta d^-_{ij}}{2}
+ \epsilon_j^2 \frac{\Delta d^+_{ij}+\Delta d^-_{ij}}{2} \, .
\end{equation*}

The theoretical prediction $t_i(\as)$ for bin $i$ depends on the value
of $\as$.
Furthermore, the theoretical predictions are also affected by sources of 
correlated uncertainties; $\delta_{ik}(\lambda_k)$ specifies
the relative uncertainty of $t_i$ due to the source $k$.
Like the $\epsilon_j$, the $\lambda_j$ are also treated as 
Gaussian distributed random variables 
with zero expectation value and unity width.
It is assumed that the theoretical predictions can be obtained
with statistical uncertainties which are negligible as compared to the
statistical uncertainties of the measurements.

The continuous dependence of the relative uncertainty $\delta_{ik}(\lambda_k)$ 
can be obtained through quadratic interpolation between 
the central result $t_i$ and the results $t^\pm_{ik}$ 
obtained by variations corresponding to plus/minus one standard deviation
due to source $k$
\begin{equation*}
 \delta_{ik}(\lambda_k)\, =\,  \lambda_j 
\frac{t^+_{ik}-t^-_{ik}}{2t_i}
\, +\, \lambda_k^2 \left( \frac{t^+_{ik}+t^-_{ik}}{2t_i} - 1 \right) \, .
\end{equation*}
The $\chi^2$ used in the $\as$ extraction is then computed as
\begin{equation*}
  \chi^2(\as,\vec{\epsilon},\vec{\lambda}) \, = \, \sum_i  
   \frac{ \left[ d_i - t_i(\as)
   \frac{\left(1 + \sum_k \delta_{ik}(\lambda_k)  \right)  }
       {\left(1+\sum_j \delta_{ij}(\epsilon_j)\right)}
         \right]^2}{
      \sigma^2_{i,\mathrm{stat.}} + \sigma^2_{i,\mathrm{uncorr.}}} 
    \, + \, \sum_j \epsilon_j^2  \, + \, \sum_k \lambda_k^2   \, ,
\label{eq:chi2main}
\end{equation*}
where $i$ runs over all data points, $j$ runs over all sources of experimental 
correlated uncertainties, and $k$ over all theoretical correlated uncertainties.
The fit result of $\as$ is determined by minimizing $\chi^2$ with respect 
to $\as$ and the ``nuisance parameters'' $\epsilon_j$ and $\lambda_k$.

\section{On the compatibility of the $\Rdphi$ data 
         and the world average of $\asmz$\label{sec:asworld}}

The $\asmz$ result in Table~\ref{tab:asmz} is lower than the world average
value by approximately one standard deviation.
In this appendix, the consistency of the world average of $\asmz$
and the $\Rdphi$ data is investigated using the $\chi^2$ values.
The $\chi^2$ values are computed according to Appendix~\ref{sec:chi2},
using the 18 data points with $\Dphimax = 7\pi/8$,
and $0.0<\ystar<0.5$ and $0.5<\ystar<1.0$.
The theoretical predictions are computed for the fixed value
of $\asmz=0.1181$.
The computation of $\chi^2$ 
uses the Hessian method for the treatment of all
uncertainties except for the PDF set uncertainty 
and the scale dependence,
so the $\chi^2$ values do not reflect these theoretical uncertainties.
Therefore, a series of $\chi^2$ values is computed for
possible combinations of variations of $\mur$ and $\muf$
around the central choice $\mur = \muf = \mu_0 = \Ht/2$.
The results are displayed in Table~\ref{tab:asworld}
and compared to the $\chi^2$ values obtained 
when $\asmz$ is a free fit parameter.

\begin{table}[t]
\centering
\caption{\label{tab:asworld}
The $\chi^2$ values between the 18 data points
and the theoretical predictions
when $\asmz$ 
is fixed to the world average value of $\asmz = 0.1181$ (third column)
and when it is a free fitted parameter (fourth column) 
for variations of the scales $\mu_R$ and $\mu_F$ around the central
choice $\mu_R = \mu_F = \mu_0 = \Ht/2$.}
\begin{tabular}{lccr}
\hline \hline
$\mu_R / \mu_0$ &  $\mu_F / \mu_0$ & $\chi^2$ for & $\chi^2$ for \\
&    &   $\asmz=$  &  $\asmz$ free \\
&   &   $0.1181$  &  fit parameter \\
\hline
0.5  &  0.5   &   62.4  & 50.9 \\
0.5  &  1.0   &   56.3  & 39.6 \\
1.0  &  0.5   &   31.6  & 23.6 \\
1.0  &  1.0   &   29.7  & 21.7  \\
1.0  &  2.0   &   28.4  & 20.8 \\
2.0  &  1.0   &   19.2  & 19.0 \\
2.0  &  2.0   &   19.3  &  19.3 \\
\hline \hline
\end{tabular}
\end{table}

When $\asmz$ is fixed to the world average, 
the $\chi^2$ value for the central scale choice
is slightly higher
than the one obtained for a free $\asmz$,
and also higher than the expectation of
$\chi^2 = N_\mathrm{dof} \pm \sqrt{2 \cdot N_\mathrm{dof}}$,
where $N_\mathrm{dof} = 18$ when $\asmz$ is fixed or $17$ 
when it is a free fit parameter.
However, the $\chi^2$ definition does not take into account 
the theoretical uncertainty due to the scale dependence.
When the renormalization scale is increased by a factor of two,
to $\mur = 2 \mu_0$, lower $\chi^2$ values are obtained,
which are similar in size to the ones obtained for a free $\asmz$,
and close to the expectation
(the dependence on the factorization scale is rather small).
Since these $\chi^2$ values are well within the range of the expectation,
it is concluded that, within their uncertainties,
the theoretical predictions for the world average value of $\asmz$
are consistent with the $\Rdphi$ data.

\section{On the compatibility of the RGE and the slope of the $\asq$ 
         results \label{sec:asslope}}

It is natural to ask whether the observed $Q$ dependence (i.e.\ the running)
of the $\asq$ results shown in Figure~\ref{fig:alphas1} 
is described by the RGE or instead exhibits significant deviations 
at the highest $Q$ values, possibly indicating signals of physics 
beyond the Standard Model.
The consistency of the RGE predictions with the observed slope
is investigated in this appendix.
The RGE prediction would be in agreement with the 
observed $Q$ dependence of the $\asq$ results
if the latter, when evolved to $\mZ$, 
give $\asmz$ values that are independent of $Q$.
For this purpose, 
a linear function in $\log_{10}(Q/1\,\mathrm{\GeV})$,
$f(Q) = c + m\cdot\log_{10}(Q/1\,\mathrm{\GeV})$,
is fitted to the nine $\asmz$ points in Figure~\ref{fig:alphas1} (bottom)
and their statistical uncertainties.
Here the correlated systematic uncertainties are not taken into account
as their correlations are non-trivial
since the individual $\asq$ results are obtained in separate fits,
with different optimizations for the nuisance parameters.
The fit results for the slope parameter $m$ and its uncertainty
are displayed in Table~\ref{tab:asslope} for a fit to the 
$\asmz$ points at all nine $Q$ values, and also for fits to different subsets 
of the $\asmz$ points, omitting points either 
at lower or higher $Q$.

\begin{table}[t]
\centering
\caption{\label{tab:asslope}Fit 
  of a linear function in $\log_{10}(Q/\mathrm{GeV})$
  to the nine extracted $\asq$ results  
  with their statistical uncertainties.
}
\begin{tabular}{lrr}
\hline \hline
 $\asq$ points  &  $Q$ range &   Fit result for  \\
 included in fit  & (\GeV) &  slope parameter \\
\hline
1--9  & $225$--$2000$  &   ($-$0.89 $\pm$ 0.35) $ \cdot 10^{-2}$ \\
2--9  & $300$--$2000$   &   ($-$0.52 $\pm$ 0.33) $ \cdot 10^{-2}$ \\
3--9  & $375$--$2000$   &   ($-$0.39 $\pm$ 0.28) $ \cdot 10^{-2}$ \\
4--9  & $450$--$2000$   &   ($-$0.20 $\pm$ 0.29) $ \cdot 10^{-2}$ \\
5--9  & $550$--$2000$   &   ($-$1.19 $\pm$ 0.35) $ \cdot 10^{-2}$ \\
6--9  & $700$--$2000$   &   ($+$0.35 $\pm$ 0.51) $ \cdot 10^{-2}$ \\
\hline
1--9  & $225$--$2000$  &   ($-$0.89 $\pm$ 0.35) $ \cdot 10^{-2}$  \\
1--8  & $225$--$1350$  &   ($-$0.85 $\pm$ 0.43) $ \cdot 10^{-2}$  \\
1--7  & $225$--$1100$  &   ($-$0.78 $\pm$ 0.32) $ \cdot 10^{-2}$  \\
1--6  & $225$--$900$  &   ($-$1.14 $\pm$ 0.28) $ \cdot 10^{-2}$  \\
1--5  & $225$--$700$  &   ($-$1.01 $\pm$ 0.31) $ \cdot 10^{-2}$  \\
1--4  & $225$--$550$  &   ($-$2.55 $\pm$ 0.41) $ \cdot 10^{-2}$  \\ 
\hline \hline
\end{tabular}
\end{table}

As documented in Table~\ref{tab:asslope},
a fit to all nine $\asmz$ points gives a slope
that differs from zero by more than its uncertainty.
Fits to groups of data points, however, show that the significance
of this slope arises from the two points at lowest $Q$.
Omitting the $\asmz$ point at lowest $Q$
(fitting points \# 2--9), 
or the two points at lowest $Q$ 
(fitting points \# 3--9), 
both give fit results for which the slope parameter is more consistent
with zero, while the $\asmz$ results change by less than $\pm 0.0001$.
On the other hand, omitting the $\asq$ points at highest $Q$
(fitting points \# 1--8 or \# 1--7) 
does not affect the significance of the slope.
It is therefore concluded that the high-$Q$ behavior of the 
$\asq$ results is consistent with the RGE
and that the small differences at lowest $Q$ do not affect 
the combined $\asmz$ result.

\printbibliography

\clearpage 
 
\begin{flushleft}
{\Large The ATLAS Collaboration}

\bigskip

M.~Aaboud$^\textrm{\scriptsize 34d}$,
G.~Aad$^\textrm{\scriptsize 99}$,
B.~Abbott$^\textrm{\scriptsize 124}$,
O.~Abdinov$^\textrm{\scriptsize 13,*}$,
B.~Abeloos$^\textrm{\scriptsize 128}$,
S.H.~Abidi$^\textrm{\scriptsize 164}$,
O.S.~AbouZeid$^\textrm{\scriptsize 143}$,
N.L.~Abraham$^\textrm{\scriptsize 153}$,
H.~Abramowicz$^\textrm{\scriptsize 158}$,
H.~Abreu$^\textrm{\scriptsize 157}$
% ...
\bigskip
\\

% ...
$^{13}$Institute of Physics, Azerbaijan Academy of Sciences, Baku; Azerbaijan.\\
% ...
$^{34}$$^{(a)}$Facult\'e des Sciences Ain Chock, R\'eseau Universitaire de Physique des Hautes Energies - Universit\'e Hassan II, Casablanca;$^{(b)}$Centre National de l'Energie des Sciences Techniques Nucleaires, Rabat;$^{(c)}$Facult\'e des Sciences Semlalia, Universit\'e Cadi Ayyad, LPHEA-Marrakech;$^{(d)}$Facult\'e des Sciences, Universit\'e Mohamed Premier and LPTPM, Oujda;$^{(e)}$Facult\'e des sciences, Universit\'e Mohammed V, Rabat; Morocco.\\
% ...
$^{99}$CPPM, Aix-Marseille Universit\'e and CNRS/IN2P3, Marseille; France.\\
% ...
$^{124}$Homer L. Dodge Department of Physics and Astronomy, University of Oklahoma, Norman OK; United States of America.\\
% ...
$^{128}$LAL, Universit\'e Paris-Sud, CNRS/IN2P3, Universit\'e Paris-Saclay, Orsay; France.\\
% ...
$^{143}$Santa Cruz Institute for Particle Physics, University of California Santa Cruz, Santa Cruz CA; United States of America.\\
% ...
$^{153}$Department of Physics and Astronomy, University of Sussex, Brighton; United Kingdom.\\
% ...
$^{157}$Department of Physics, Technion: Israel Institute of Technology, Haifa; Israel.\\
$^{158}$Raymond and Beverly Sackler School of Physics and Astronomy, Tel Aviv University, Tel Aviv; Israel.\\
% ...
$^{164}$Department of Physics, University of Toronto, Toronto ON; Canada.\\
% ...
$^{a}$ Also at Borough of Manhattan Community College, City University of New York, New York City; United States of America.\\
$^{b}$ Also at Centre for High Performance Computing, CSIR Campus, Rosebank, Cape Town; South Africa.\\
$^{c}$ Also at CERN, Geneva; Switzerland.\\
$^{d}$ Also at CPPM, Aix-Marseille Universit\'e and CNRS/IN2P3, Marseille; France.\\
$^{e}$ Also at Departament de Fisica de la Universitat Autonoma de Barcelona, Barcelona; Spain.\\
% ...
$^{*}$ Deceased

\end{flushleft}


\end{document}